\documentclass{article}
\usepackage[T1]{fontenc}
\usepackage[utf8]{inputenc}
\usepackage{lmodern}
\usepackage[english]{babel}
\usepackage{amsmath}
\usepackage{amsfonts}
\usepackage{mathrsfs}
\usepackage{amsthm}
\usepackage{amssymb}
\usepackage{latexsym}
\usepackage{enumerate}
\usepackage{graphicx}
\usepackage{bbold}
\usepackage{subfigure}
\usepackage{epstopdf}
\usepackage[affil-it,auth-sc]{authblk}
\usepackage{cite}

\begin{document}

\title{Three-sphere low-Reynolds-number swimmer with muscle-like arms}

\author[1]{Alessandro Montino}
\author[1,2,\footnote{Corresponding author -- e-mail address:\,\texttt{desimone@sissa.it} -- phone:\,+39\,040\,3787\,455}]{Antonio DeSimone}
\affil[1]{\small{GSSI - Gran Sasso Science Institute, viale Francesco Crispi 7, 67100 L'Aquila, Italy.} \smallskip}
\affil[2]{\small{SISSA - International School for Advanced Studies, via Bonomea 265, 34136 Trieste, Italy.} \smallskip}
\maketitle

\begin{abstract}
The three-sphere swimmer by Najafi and Golestanian is composed of three spheres connected by two arms. The authors of this model studied in detail the case in which the swimmer can generate periodic shape changes by controlling the lengths of the two arms. Here we study a variation of the model in which the geometry of the shape change is not known a priori, because the swimmer is not able of directly controlling the lengths of the arms. Our study is motivated by the fact that real swimmers are not capable of directly contolling their shape. The arms of our three-sphere swimmer are constructed according to Hill's model of muscular contraction. The swimmer is only able to control the forces developed in the active components of the muscle-like arms. The two shape parameters and the forces acting through the two arms evolve according to a system of ODEs. After giving a mathematical formulation of the problem, we study the qualitative properties of the solutions and compute analytically their leading order approximation. Then we present the results of some numerical simulations which are in good agreement with our theoretical predictions. Finally, we study some optimization problems. Our results can help to gain insight into the mechanisms governing locomotion of biological swimmers.
\end{abstract}

\section*{Introduction}

The three-sphere swimmer by Najafi and Golestanian \cite{Najafi-Golestanian, golestanian2008analytic} is a cornerstone in the literature on low Reynolds number swimming. It is composed of three spheres connected by two arms. The presence of two shape parameters (the lengths of the two arms) allows to perform periodic shape changes which are not invariant under time reversal. This is the key to beat Purcell's famous \textit{scallop theorem} \cite{Purcell}, which says that low Reynolds number swimmers cannot achieve locomotion through a reciprocal shape change. Another famous example of low Reynolds number swimmer with two shape parameters is the three-link swimmer by Purcell \cite{Purcell}. These two model swimmers have been studied extensively (see for example \cite{BKS03,AKO05,AGK04,TH07,ADSGZ13,ADSL08,ADSL09,ADSH11,ADSHLM13,DSHAL12}). 

In the works \cite{Najafi-Golestanian,golestanian2008analytic} the authors study a three-sphere swimmer which is able to perform periodic shape changes by controlling the lengths of the two arms. Real swimmers however are not capable of directly controlling their shape. Their shape change is generated by a complex interplay between different elements: forces generated by the swimmer through internal mechanisms, elastic properties of the swimmer's body, and forces due to interactions with the surrounding viscous fluid. The swimmer is able to control only the first of these components. For this reason it interesting to study minimal swimmers which are not capable of directly controlling their shape. An interesting work going in this direction is \cite{GK2008}, in which the authors consider a three-sphere swimmer whose arms can generate forces thanks to a system of molecular motors and elastic elements. Another possibility is to assume that one of the two shape parameters can be controlled while the other one is driven by a passive elastic spring. This was done in \cite{Or} for the three-link swimmer and in \cite{MDS2015} for the three-sphere swimmer. 

In the present work we study a three-sphere swimmer whose arms are viscoelastic structures constructed according to Hill's active state muscle model. This model, which is very useful for computing the mechanical behaviour of muscles, includes passive elastic elements, a mechanism of viscous damping, and an active component. The swimmer is able to control the force generated by the active component. The geometry of the shape change is not known a priori and will have to be computed by  solving a system of ODEs. In the first section we introduce our model and derive the ODEs governing the system. In the second section we study the qualitative properties of solutions, compute analytically the leading order term of their asymptotic expansion, present the results of numerical simulations, and study some optimmization problems. In the third section we study a variation of our model, in which one of the two muscle-like arms is replaced by a passive elastic spring.


\section{Problem formulation}
The aim of this section is to introduce the thee-sphere swimmer with muscle-like arms and to obtain a system of ODEs governing its dynamics. In the first subsection we fix the notations and recall some basic facts about the three-sphere swimmer. In the second subsection we assume that the forces acting through the two arms are known functions of time and shape and obtain the equations governing the evolution of the two shape parameters. In the third subsection we model the arms according to Hill's muscle model and obtain a system of ODEs governing  the forces acting across each arm, the shape of the swimmer and its displacement. In the fourth subsection we write the problem in non-dimensional form. In the fifth and last subsection we consider a variation of the problem in which one of the two muscle-like arms is replaced by a passive elastic spring.

\subsection{Three-sphere swimmer}

\setlength{\unitlength}{1cm}
\begin{figure}[h!]
\begin{picture}(10,5)(-5,-2.5) 
\put(0,0){\circle{1}}
\put(-4,0){\circle{1}}
\put(6,0){\circle{1}}
\thicklines
\put(7,0){\vector(1,0){0.75}}
\thinlines
\put(7.15,0.25){$\mathbf{e}$}
\put(.5,0){\line(1,0){5}}
\put(-0.5,0){\line(-1,0){3}}
\multiput(-4,0)(0,-0.2){8}{\line(0,-1){0.1}}
\multiput(0,0)(0,-0.2){8}{\line(0,-1){0.1}}
\multiput(6,0)(0,-0.2){8}{\line(0,-1){0.1}}
\put(-2.5,-1){\vector(-1,0){1.5}}
\put(-1.5,-1){\vector(1,0){1.5}}
\put(-2.2,-1.2){$L_1$}
\put(2.5,-1){\vector(-1,0){2.5}}
\put(3.5,-1){\vector(1,0){2.5}}
\put(2.8,-1.2){$L_2$}
\put(0,0){\vector(-1,1){0.36}}
\put(-0.2,0.2){$a$}
\put(-4,1){$1$}
\put(0,1){$2$}
\put(6,1){$3$}
\end{picture}
\caption{\label{TSS} The three-sphere low Reynolds number swimmer.}
\end{figure}
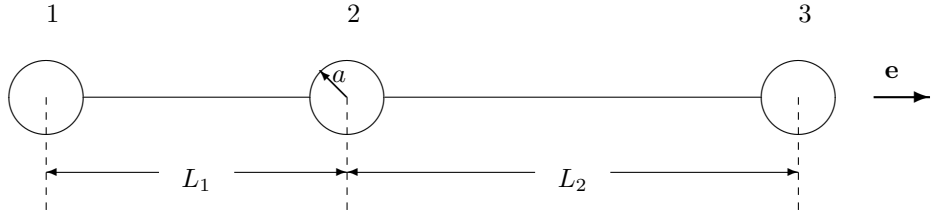
Let us consider a three-sphere swimmer as in Figure \ref{TSS}. Let $a$ be the radius of the spheres, and $L_1, L_2$ the lengths of the two arms. We indicate with $\mu$ the dynamic viscosity of the fluid and with $v_i$ and $f_i$ respectively the velocity of sphere $i$ and the force that this sphere exerts on the fluid, projected on the unit vector $\mathbf{e}$. By using the Oseen tensor and the approximation $\frac{a}{L_i}<<1$ we obtain the following linear relations between forces and velocities
\begin{align}
v_1&=\frac{f_1}{6 \pi \mu a}+\frac{f_2}{4 \pi \mu L_1}+\frac{f_3}{4 \pi \mu (L_1+L_2)} \label{oseen1} \,, \\
v_2&=\frac{f_1}{4 \pi \mu L_1}+\frac{f_2}{6 \pi \mu a}+\frac{f_3}{4 \pi \mu L_2} \label{oseen2} \,, \\
v_3&=\frac{f_1}{4 \pi \mu (L_1+L_2)}+\frac{f_2}{4 \pi \mu L_2}+\frac{f_3}{6 \pi \mu a} \label{oseen3} \,.
\end{align}
Due to Newton's third law of motion, the force exerted by the fluid on the $i$-th sphere is $-f_i$. Therefore the force balance equation for the swimmer is
\begin{equation} \label{GFB}
f_1+f_2+f_3=0 \,.
\end{equation}
The geometry of the system implies the following kinematic relations
\begin{align}
\dot{L}_1=v_2-v_1  \label{kine1}\,,\\ 
\dot{L}_2=v_3-v_2  \label{kine2} \,.
\end{align}

Let us indicate with $x_i$ the position of the $i$-th sphere on axis corresponding to the unit vector $\mathbf{e}$. We indicate with $x$ the mean point of the three spheres, namely, 
\begin{equation}
x:=\frac{1}{3}(x_1+x_2+x_3) \,.
\end{equation}
The translational velocity of the swimmer is the velocity of the point $x$. Obviously
\begin{equation}
\dot x=\frac{1}{3}(v_1+v_2+v_3) \,.
\end{equation}
Using equations \eqref{oseen1}-\eqref{GFB} we can show that
\begin{equation} \label{golestanian6}
\dot x=\Big( \frac{1}{L_1+L_2}-\frac{1}{L_2} \Big)\frac{f_1}{12 \pi \mu}+\Big( \frac{1}{L_1+L_2}-\frac{1}{L_1}\Big)\frac{f_3}{12 \pi \mu} \,.
\end{equation}
Now we can use equations \eqref{oseen1}-\eqref{kine2} to express $f_1$ and $f_3$ in terms of $L_1,L_2,\dot{L}_1,\dot{L}_2$. If we plug the resulting expressions into equation \eqref{golestanian6} and keep only the leading order terms in $a/L_j$ we obtain
\begin{equation} \label{V_0}
\dot x=\frac{a}{6}\Big[ \Big( \frac{\dot{L}_2-\dot{L}_1}{L_2+L_1} \Big)+2\Big( \frac{\dot{L}_1}{L_2}-\frac{\dot{L}_2}{L_1}\Big)+\frac{\dot{L}_2}{L_2}-\frac{\dot{L}_1}{L_1} \Big]\,.
\end{equation}
Suppose that $L_1$ and $L_2$ are periodic functions. In this case the terms $\dot{L}_j/L_j$ average to zero in  a full swimming cycle, because they are derivatives of $\log(L_j)$. If we neglect these terms we obtain the following formula
\begin{equation} \label{V_0_bis}
\dot x=\frac{a}{6}\Big[ \Big( \frac{\dot{L}_2-\dot{L}_1}{L_2+L_1} \Big)+2\Big( \frac{\dot{L}_1}{L_2}-\frac{\dot{L}_2}{L_1}\Big) \Big]\,,
\end{equation}
which  can be used instead of \eqref{V_0} to compute the net displacement in one period when the swimmer performs a periodic shape change.

Now we consider the case of small deformations
\begin{equation} \label{small_def}
\begin{cases}
&L_1=l_1+U_1 \\
&L_2=l_2+U_2 \\
&U_i/l_j<<1 \,.
\end{cases}
\end{equation}
We assume that the deformations are periodic with period $T$. We would like to compute the net displacement in one period to leading order in the amplitude of deformations. Let us set $l:=l_1+l_2$. $l$ is the body length of the swimmer. If we expand equation \eqref{V_0_bis} and retain only the leading order terms in $U_i/l_j$ we get
\begin{equation}
\dot \xi=\frac{a}{6} \Big[ \frac{\dot{U_2}-\dot{U_1}}{l_1+l_2}-\frac{(\dot{U_2}-\dot{U_1})(U_1+U_2)}{(l_1+l_2)^2} +\frac{2 \dot{U_1}}{l_2}-\frac{2 \dot{U_1} U_2}{l_2^2}-\frac{2 \dot{U_2}}{l_1}+\frac{2\dot{U_2}U_1}{l_1^2} \Big] \,.
\end{equation}
The terms $\dot{U_j},\dot{U_j}U_j,\dot{U_1}U_2+U_1\dot{U_2}$ give zero when integrated between $0$ and $T$. So the net displacement in one period $l \Delta x$ is given, to leading order in $U_i/l_j$, by the following formula
\begin{equation} \label{deltax}
l \Delta x= \frac{a }{6} \Big[ \frac{1}{l_1^2}+\frac{1}{l_2^2}-\frac{1}{(l_1+l_2)^2} \Big] \int_0^T(U_1 \dot{U_2}-\dot{U_1} U_2) dt \,.
\end{equation}

\subsection{TSS with assigned tensions}

Let us indicate with $T_1$ and $T_2$ the tension on the tail rod and on the front rod respectively. We assume that $T_1$ and $T_2$ are known functions of time and shape, namely, $T_i=T_i(t,L_1,L_2)$, $i=1,2$. The equations of force balance for the three spheres are
\begin{align}
&-f_1+T_1=0  \label{balance1}\\
&-f_2-T_1+T_2=0  \label{balance2} \\
&-f_3-T_2=0 \label{balance3} \,.
\end{align}
Notice that equations \eqref{balance1}-\eqref{balance3} imply that the condition of global force balance \eqref{GFB} is satisfied. Our aim now is to obtain two ODEs for $L_1$ and $L_2$. From equations \eqref{balance1}-\eqref{balance3} we can easily express the forces as functions of $T_1$ and $T_2$. If we plug the resulting expressions into equations \eqref{oseen1}-\eqref{oseen3} we obtain
\begin{align}
v_1&=\frac{T_1}{6 \pi \mu a}+\frac{T_2-T_1}{4 \pi \mu L_1}-\frac{T_2}{4 \pi \mu (L_1+L_2)} \,, \\
v_2&=\frac{T_1}{4 \pi \mu L_1}+\frac{T_2-T_1}{6 \pi \mu a}-\frac{T_2}{4 \pi \mu L_2} \,, \\
v_3&=\frac{T_1}{4 \pi \mu (L_1+L_2)}+\frac{T_2-T_1}{4 \pi \mu L_2}-\frac{T_2}{6 \pi \mu a}  \,.
\end{align}
These equations, combined with the kinematic relations \eqref{kine1} and \eqref{kine2}, yield the following system of ODEs governing the evolution of $L_1$ and $L_2$
\begin{align}
\dot{L}_1&=\frac{1}{\pi \mu}\Big(\frac{1}{2 L_1}-\frac{1}{3a}\Big)T_1+\frac{1}{\pi \mu} \Big(\frac{1}{6a}-\frac{1}{4 L_1}-\frac{1}{4 L_2}+\frac{1}{4(L_1+L_2)}\Big)T_2 \label{ODE1} \\
\dot{L}_2&=\frac{1}{\pi \mu}\Big(\frac{1}{6a}-\frac{1}{4 L_1}-\frac{1}{4 L_2}+\frac{1}{4(L_1+L_2)}\Big)T_1+\frac{1}{\pi \mu}\Big(\frac{1}{2 L_2}-\frac{1}{3a}\Big)T_2\label{ODE2} \,.
\end{align}

\subsection{TSS with muscle-like arms}

\begin{figure}[h!] 
\includegraphics[width=0.9 \textwidth]{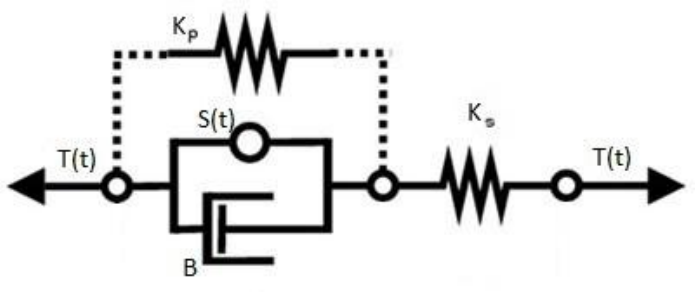}
\caption{\label{hillsmodel} Hill's active state muscle model.}
\end{figure}

Equations \eqref{ODE1} and \eqref{ODE2} drive the evolution of $L_1$ and $L_2$ once $T_1$ and $T_2$ are known functions of shape and time. In this section we introduce a further element: we model the two arms of the swimmer according to Hill's active state muscle model. A schematicrepresentation of this model is shown in Figure \ref{hillsmodel}. The contractile element is composed of an active component capable of generating a tension $S(t)$ and a linear dashpot with characterisic constant $B$. In addition there are two linear springs, one in parallel and one in series with the contractile element, with elastic constants $k_p$ and $k_s$ respectively. 
We assume that the active components of the two arms can generate tensions $S_1(t)$ and $S_2(t)$. Let us indicate with $l_i$ the sum of the rest lengths of the parallel and series springs in arm $i$ ($i=1,2$). Tensions and lengths are related by the following equations
\begin{align}
T_1&=S_1+B_1 \dot L_1-\frac{B_1}{k_{s,1}}\dot T_1+k_{p,1}(L_1-l_1)-\frac{k_{p,1}}{k_{s,1}} T_1 \\
T_2&=S_2+B_2 \dot L_2-\frac{B}{k_{s,2}}\dot T_2+k_{p,2}(L_2-l_2)-\frac{k_{p,2}}{k_{s,2}} T_2 \,.
\end{align}
By rearranging the terms we find
\begin{align}
\dot T_1&=-\frac{k_{s,1}+k_{p,1}}{B_1} T_1+\frac{k_{s,1}}{B_1} S_1+\frac{k_{s,1} k_{p,1}}{B_1}(L_1-l_1)+k_{s,1} \dot L_1 \label{ODE3} \\
\dot T_2&=-\frac{k_{s,2}+k_{p,2}}{B_2} T_2+\frac{k_{s,2}}{B_2} S_2+\frac{k_{s,2} k_{p,2}}{B_2}(L_2-l_2)+k_{s,2} \dot L_2 \label{ODE4} \,.
\end{align}
By using equations \eqref{ODE1} and \eqref{ODE2} we can express $\dot L_1$ and $\dot L_2$ as functions of $T_1$, $T_2$, $L_1$, and $L_2$. Then we can plug the resulting expressions into \eqref{V_0}, \eqref{ODE3}, and \eqref{ODE4}. As a consequence, we can rewrite \eqref{ODE1}-\eqref{ODE4} and \eqref{V_0} as a system of five ODEs
governing the evolution of $x,L_1,L_2,T_1$, and $T_2$.

\subsection{Non-dimensionalization}

The aim of this subsection is to write the problem in non-dimensional form. We will discover that the physical parameters influence the behaviour of the dynamics through $6$ non-dimensional numbers. This approach allows to obtain more general and useful results.  Let $l=l_1+l_2$ be the characteristic length of the swimmer and $1/\omega$ a characteristic time. Let us set $t^*= \omega t$, $\Lambda_i=L_i/l$, $\lambda_i=l_i/l$, $\xi=x/l$, $\alpha=a/l$, $\tau_i=T_i/(\mu \omega l^2)$, $\sigma_i=S_i/(\mu \omega l^2)$. We will indicate with the prime symbol the derivative with respect to the rescaled time $t^*$. Using the new variables our system can be rewritten as
\begin{align}
\Lambda_1'&=\frac{1}{\pi} \Big(\frac{1}{2 \Lambda_1}-\frac{1}{3 \alpha} \Big) \tau_1+\frac{1}{\pi} \Big(\frac{1}{6 \alpha}-\frac{1}{4 \Lambda_1}-\frac{1}{4 \Lambda_2}+\frac{1}{4(\Lambda_1+\Lambda_2)} \Big) \tau_2 \label{ODEsystem1} \\
\Lambda_2'&=\frac{1}{\pi}  \Big(\frac{1}{6 \alpha}-\frac{1}{4 \Lambda_1}-\frac{1}{4 \Lambda_2}+\frac{1}{4(\Lambda_1+\Lambda_2)} \Big) \tau_1+\frac{1}{\pi} \Big( \frac{1}{2 \Lambda_2}-\frac{1}{3 \alpha} \Big) \tau_2 \label{ODEsystem2}\\
\xi '&=\frac{\alpha}{6} \Big(-\frac{1}{\Lambda_1+\Lambda_2}+\frac{2}{\Lambda_2}-\frac{1}{\Lambda_1} \Big) \Lambda_1'+\frac{\alpha}{6} \Big( \frac{1}{\Lambda_1+\Lambda_2}-\frac{2}{\Lambda_1}+\frac{1}{\Lambda_2} \Big) \Lambda_2' \label{ODEsystem3}\\
\tau_1'&=-(J_{s,1}+J_{p,1}) \tau_1+J_{s,1} \sigma_1+J_{p,1}K_1 (\Lambda_1-\lambda_1)+K_1 \Lambda_1' \label{ODEsystem4} \\
\tau_2'&=-(J_{s,2}+J_{p,2}) \tau_2+J_{s,2} \sigma_2+J_{p,2}K_2 (\Lambda_2-\lambda_2)+K_2 \Lambda_2' \label{ODEsystem5}  \,,
\end{align}
where the dimensionless parameters are
\begin{equation}
J_{s,i} :=\frac{k_{s,i}}{\omega B_i},\ \ 
J_{p,i}:=\frac{k_{p,i}}{\omega B_i},\ \ 
K_i := \frac{k_{s,i}}{\mu \omega l} \,,
\end{equation}
for $i=1,2$.

\subsection{TSS with one muscle-like arm and one passive elastic arm}

Now we replace the tail arm with a passive elastic spring. Let us indicate with $l_1$ the rest length and with $h$ the elastic constant of the spring. Equations \eqref{ODEsystem1}-\eqref{ODEsystem3} and \eqref{ODEsystem5} remain valid. Equation \eqref{ODEsystem4} is replaced by
\begin{equation} \label{tau1_elastic}
\tau_1=H (\Lambda_1-\lambda_1) \,,
\end{equation}
where $\lambda_1=l_1/l$ and $H=h/(\omega \mu l)$.

\section{TSS with two muscle-like arms}

In this section we present a detailed study of the TSS with two muscle-like arms.  For the sake of simplicity we assume that the two arms are identical. Therefore
\begin{align}
&\lambda_1=\lambda_2=1/2 \\
&K_1=K_2=:K \\
&J_{s,1}=J_{s,2}=:J_{s} \\
&J_{p,1}=J_{p,2}=:J_{p} \,.
\end{align}
We consider the case in which the tensions developed by the active components are small, namely,
\begin{equation}
\sigma_i=\epsilon \tilde \sigma_i \,,
\end{equation}
with $\epsilon <<1$ and $|\tilde \sigma_i| \le 1$. We assume that $\sigma_1$ and $\sigma_2$ are periodic. In the first subsection we study the qualitative properties of the solutions. In the second subsection we compute the solutions analytically to leading order in $\epsilon$. In the third subsection we present the results of numerical simulations and in the fourth one we study some optimization problems.

\subsection{Qualitative properties of the solutions} \label{qualitative}

In this subsection we prove that, for small enough values of $\epsilon$, there exists one and only one periodic orbit and that this orbit is asymptotically stable. Our analysis is based on some classical results on periodically perturbed systems which can be found in \cite{farkas1994periodic}.

We observe that in the ODEs \eqref{ODEsystem1}-\eqref{ODEsystem5}, $\xi$ does not appear in the velocity field. This is due to the fact that the response of the system is invariant under translations. As a consequence, the differential problem for $\xi$ is decoupled from the rest. So we can restrict our attention to the four-dimensional system for $\Lambda_1,\Lambda_2,\tau_1,\tau_2$. Instead of $\Lambda_1$ and $\Lambda_2$ we use the variables $u_1=\Lambda_1-\lambda_1$ and $u_2=\Lambda_2-\lambda_2$. Let us set $Y:=(u_1,u_2,\tau_1,\tau_2)^{tr}$ and $\tilde J:=J_s+J_p$. We can write our system in the form
\begin{equation}
Y '=f(Y)+\epsilon g(t^*) \,,
\end{equation}
where $f(Y)$ is
\begin{equation}
\left( \begin{array}{c}  \frac{1}{\pi} \Big(\frac{1}{2(\lambda_1+u_1)}-\frac{1}{3 \alpha} \Big) \tau_1+\frac{1}{\pi} \Big(\frac{1}{6 \alpha}-\frac{1}{4(\lambda_1+u_1)}-\frac{1}{4(\lambda_2+ u_2)}+\frac{1}{4(\lambda_1+u_1+\lambda_2+u_2)} \Big) \tau_2\\ \frac{1}{\pi} \Big(\frac{1}{6 \alpha}-\frac{1}{4(\lambda_1+u_1)}-\frac{1}{4(\lambda_2+ u_2)}+\frac{1}{4(\lambda_1+u_1+\lambda_2+u_2)} \Big) \tau_1+\frac{1}{\pi}\Big(\frac{1}{2(\lambda_2+u_2)}-\frac{1}{3 \alpha} \Big) \tau_2  \\ J_p K u_1+[\frac{K}{\pi} \Big(\frac{1}{2(\lambda_1+u_1)}-\frac{1}{3 \alpha} \Big) -\tilde J]\tau_1+ \frac{K}{\pi} \Big(\frac{1}{6 \alpha}-\frac{1}{4(\lambda_1+u_1)}-\frac{1}{4(\lambda_2+ u_2)}+\frac{1}{4(\lambda_1+u_1+\lambda_2+u_2)} \Big) \tau_2\\  J_p K u_2 +[\frac{K}{\pi}\Big(\frac{1}{2(\lambda_2+u_2)}-\frac{1}{3 \alpha} \Big)-\tilde J ] \tau_2+\frac{K}{\pi}  \Big(\frac{1}{6 \alpha}-\frac{1}{4(\lambda_1+u_1)}-\frac{1}{4(\lambda_2+ u_2)}+\frac{1}{4(\lambda_1+u_1+\lambda_2+u_2)} \Big) \tau_1 \end{array} \right) \,,
\end{equation}
and
\begin{equation} \label{g}
g(t^*)=\left(\begin{array}{c} 0 \\ 0 \\ J_s \tilde \sigma_1(t^*) \\ J_s \tilde \sigma_2(t^*) \end{array} \right) \,.
\end{equation}
The unperturbed system $Y'=f(Y)$ has the equilibrium point $Y=0$. The variational system with respect to this equilibrium point (see \cite{farkas1994periodic}, page 303) is
\begin{equation} \label{variational_system}
Z'=A Z \,,
\end{equation}
where
\begin{equation} \label{A}
A:=
\begin{pmatrix}
0& \ 0& \ Q& \ R \\
0& \ 0& \ R& \ Q \\
J_{p} K& \ 0& \ KQ-\tilde J& \ K R \\
0& \ J_p K& \ KR& \ KQ-\tilde J
\end{pmatrix} \,,
\end{equation}
and
\begin{align}
Q&:= \frac{1}{\pi} \Big(1-\frac{1}{3 \alpha} \Big)\\
R&:= \frac{1}{\pi} \Big( \frac{1}{6 \alpha} -\frac{3}{4} \Big) \,.
\end{align}
Now we would like to prove that all the eigenvalues of $A$ have a negative real part. If this is true then all the characteristic multipliers of system \eqref{variational_system} are in modulus strictly less than one. As a consequence we can apply theorems 6.1.1 and 6.1.3 of \cite{farkas1994periodic} and conclude that, for small enough values of $\epsilon$, there exists one and only one periodic orbit. Moreover, this orbit is asymptotically stable.
The characteristic polynomial of $A$ is
\begin{align} \label{characteristic_poly}
p_A(x)=x^4+2(\tilde J-KQ) x^3+[K^2(Q^2-R^2)+\tilde J^2-2 \tilde J K Q -2 J_{p}K Q] x^2+... \nonumber \\
...+[J_{p}K^2(Q^2-R^2)-2 J_{p}KQ \tilde J]x+(J_{p}K)^2(Q^2-R^2) \,.
\end{align}
Let us indicate with $a_j$, $j=0,1,2,3$ the coefficients, so that
\begin{equation}
p_A(x)=x^4+a_3 x^3+a_2 x^2+ a_1 x + a_0 \,.
\end{equation}
First of all we observe that $a_j >0$ for every $j$. This follows from the fact that, if $\alpha$ is small enough, then $Q<0$, $R>0$, $|Q|<R$. Having all the coefficients with the same sign is a necessary condition for $p_A(x)$ to have only roots with negative real part. The assumption that $\alpha$ is small is a natural one, because all of our analysis is based on the hypothesis that the radius of the spheres is small compared to the lengths of the arms. 
The table associated to the polynomial by means of the Routh method is
\begin{equation} \label{table_Routh_1}
\begin{bmatrix}
1& \ a_2& \ a_0 \\
a_3& \ a_1& \ 0 \\
b_3& \ b_2& \ 0 \\
c_2& \ 0& \ 0 \\
...& \ ...& \ ...
\end{bmatrix} \,,
\end{equation}
where
\begin{align}
b_3&=(a_3 a_2 -a_1)/a_3 \\
b_2&=a_0 \\
c_2&=(a_1 b_3-a_3 b_2)/b_3 \,.
\end{align}
Let us study the signs of $b_3$ and $c_2$.
Notice that
\begin{align}
b_3&=a_2-a_1/a_3 \\
&=a_2+\frac{2 J_{p}K Q \tilde J-J_{p}(KQ)^2+J_{p}(K R)^2}{2(\tilde J-KQ) } \\
&=a_2+\frac{J_{p}KQ}{2}+\frac{J_{p}K Q (\tilde J-(KR)^2/(KQ))}{2(\tilde J- KQ)} \,.
\end{align}
Now we notice that, since $|Q|<R$, we have
\begin{equation}
\tilde J-(KR)^2/(KQ) \le \tilde J-KQ \,.
\end{equation}
As a consequence,
\begin{equation} \label{lower_bound_b3}
b_3 \ge a_2+ J_{p} K Q =K^2(Q^2-R^2)+ \tilde J^2-3 J_{p} K Q-2J_{s} K Q >0 \,.
\end{equation}
Notice that $\text{sign}(c_2)=\text{sign}(a_1 b_3-a_3 b_2)$. By using \eqref{lower_bound_b3} we see that
\begin{align}
a_1 b_3-a_3 b_2 & \ge (Q^2-R^2)[2 J_{p} K^4 Q^2 +2 J_{p} \tilde J^2 K^2-4 J_{p} J_{s} K^3 Q -6 J_{p}^2 K^3 \tilde J Q] +... \nonumber \\
&...-2 J_{p}\tilde J K^3 Q(Q^2-R^2)-2 J_{p} \tilde J^3 K Q+4 (J_{p} K)^2\tilde J Q^2 +... \nonumber \\
&...+4 J_{p} J_{s} \tilde J (K Q)^2+2(\tilde J-KQ)(J_{p} K)^2 R^2 > 0 \,.
\end{align}
So $b_3$ and $c_2$ are both positive. Therefore we see from table \eqref{table_Routh_1} that there are three roots with negative real part. Since the determinant of $A$ is positive, also the fourth root must have negative real part.

\subsection{Asymptotic expansions} \label{asymptotic}

\begin{figure}[!h]
\includegraphics[width=1 \textwidth]{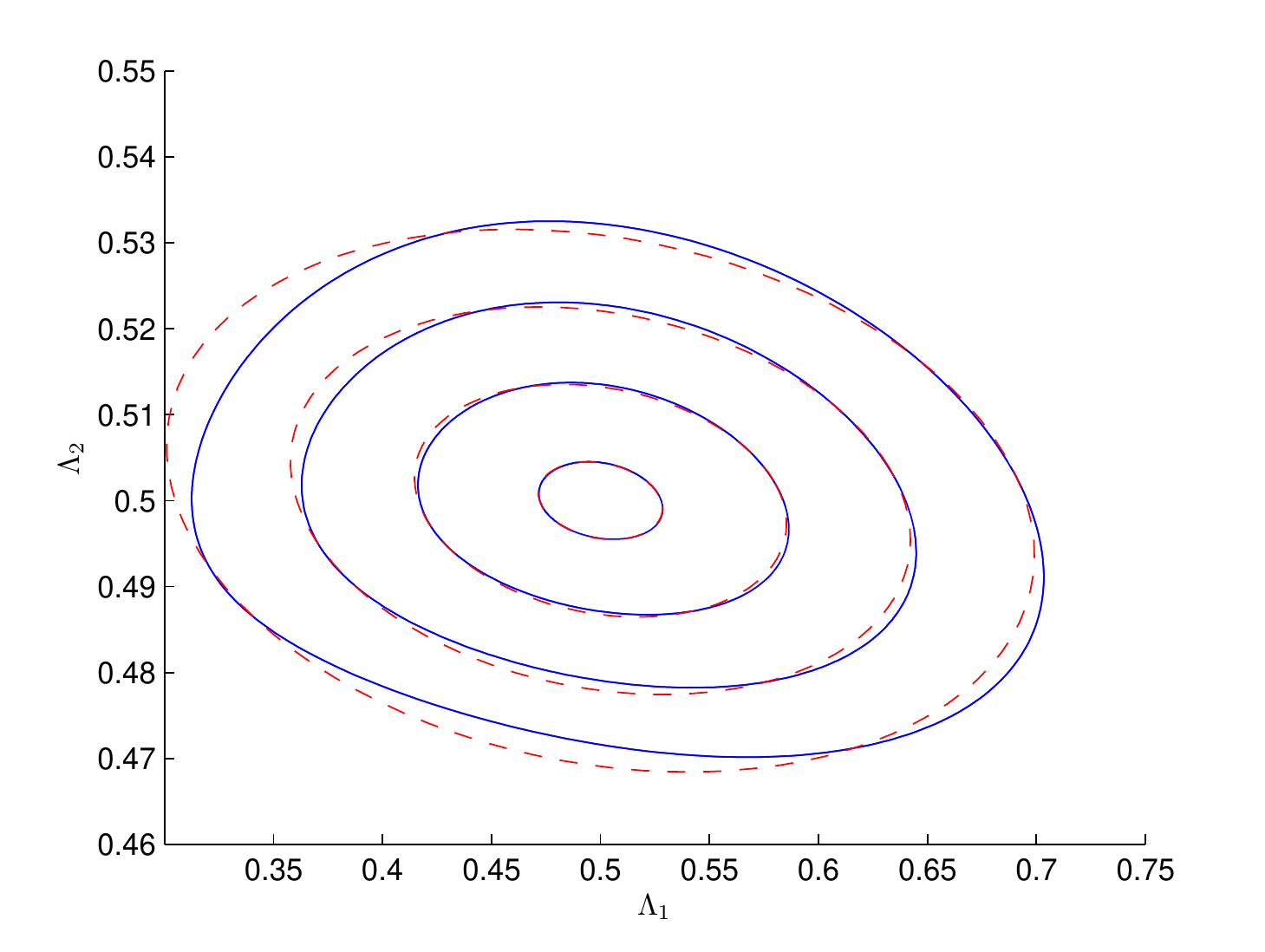}
\caption{These curves in the configuation space are the loops along which the shape of the TSS with muscle-like arms evolves. The full blue lines show the results of numerical simulations while the red dashed lines correspond to leading order approximations. The loops in this picture are obtained by varying $\epsilon$ and choosing the other parameters as in Table \ref{parameters}. The different loops, from the smallest to the largest, correspond to $\epsilon=0.1$, $\epsilon=0.3$, $\epsilon=0.5$, and $\epsilon=0.7$. We see from the picture that the error becomes vanishingly small for small values of $\epsilon$.}
\label{loops_comparison}
\end{figure}

In this section we compute analytic expressions for the solutions, to leading order in $\epsilon$.
Let us assume that
\begin{align}
\tilde \sigma_1 &= \tilde a_1 \sin (t^*)+\tilde b_1 \cos(t^*) \label{sigmatilde1} \\
\tilde \sigma_2&= \tilde a_2 \sin(t^*)+\tilde b_2 \cos(t^*)  \label{sigmatilde2} \,,
\end{align}
with $|\tilde a_j|+|\tilde b_j| \le 1$, $j=1,2$. We express the solutions in power series as follows
\begin{align}
u_1&=\epsilon u_1^{(1)}+\epsilon^2 u^{(2)}_1+... \\
u_2&=\epsilon u_2^{(1)}+\epsilon^2 u_2^{(2)}+... \\
\tau_1&=\epsilon \tau_1^{(1)}+\epsilon^2 \tau_1^{(2)}+... \\
\tau_2&=\epsilon \tau_2^{(1)}+\epsilon^2 \tau_2^{(2)}+...
\end{align}
Let us set $X=(u_1^{(1)},u_2^{(1)},\tau_1^{(1)},\tau_2^{(1)})^{tr}$. The problem satisfied by $X$ is
\begin{equation} \label{cauchy_problem}
\begin{cases}
&  X'=A X+g(t^*) \\
& X(0)=0 \,.
\end{cases}
\end{equation}
The solution of this problem can be computed with the help of Duhamel's formula
\begin{equation}
X(t^*)=\int_0^{t^*}e^{A(t^*-s)} g(s) ds \,.
\end{equation}
However, since the exponential of $A$ is difficult to compute, we choose a different strategy. We know from the previous subsection that the solution converges to a periodic orbit and we would like to study the stationary regime of the system. So we are not really interested in the solution of \eqref{cauchy_problem}. Rather, we would like to compute the periodic orbit. This problem can be reduced to the problem of solving a linear system. First of all we notice that $g$ can be rewritten as
\begin{equation}
g(t^*)=\hat g_s \sin(t^*)+\hat g_c \cos(t^*) \,,
\end{equation}
where $\hat g_s=(0,0, J_{s} \tilde a_1, J_{s}\tilde a_2)^{tr}$ and $\hat g_c=(0,0,J_{s} \tilde b_1, J_{s} \tilde b_2)^{tr}$. We look for a solution in the following form
\begin{equation}
X(t^*)= \hat X_s\sin(t^*)+\hat X_c \cos(t^*) \,,
\end{equation}
with $\hat X_s,\hat X_c \in \mathbb{R}^4$. Notice that $X'=-\hat X_c \sin(t^*)+\hat X_s \cos(t^*)$. So our differential equation becomes
\begin{equation}
-\hat X_c \sin(t^*)+\hat X_s \cos(t^*)=(A \hat X_s+\hat g_s) \sin(t^*)+(A \hat X_c+\hat g_c) \cos(t^*) \,.
\end{equation}
Since the equality must hold for every time $t^*$, we obtain the following linear system
\begin{equation}
\begin{cases}
&-\hat X_c=A \hat X_s+\hat g_s \\
&\hat X_s=A \hat X_c+ \hat g_c \,.
\end{cases}
\end{equation}
We can rewrite the problem in compact form
\begin{equation} \label{linear_system_compact}
\begin{pmatrix}
-A& \ -\mathbb{1}& \\
\mathbb{1}& \ -A \\
\end{pmatrix} \left( \begin{array}{c} \hat X_s \\ \hat X_c \end{array} \right)= \left( \begin{array}{c} \hat g_s \\ \hat g_c \end{array} \right) \,.
\end{equation}
Now suppose that
\begin{align}
u_1^{(1)}&=a_1 \sin(t^*) +b_1 \cos(t^*) \\
u_2^{(1)}&= a_2 \sin(t^*)+b_2 \cos(t^*) \\
\tau_1^{(1)}&= c_1 \sin(t^*)+d_1 \cos(t^*) \\
\tau_2^{(1)}&= c_2 \sin(t^*)+d_2 \cos(t^*) \,,
\end{align}
namely, $\hat X_s=(a_1,a_2,c_1,c_2)^{tr}$ and $\hat X_c=(b_1,b_2,d_1,d_2)^{tr}$.
The linear problem \eqref{linear_system_compact} corresponds to the following set of equations
\begin{align}
&a_1=Q d_1+R d_2 \label{linear_sys_1} \\
&a_2=R d_1+Q d_2  \label{linear_sys_2}\\
&b_1=-Qc_1-Rc_2 \label{linear_sys_3} \\
&b_2=-Rc_1-Q c_2 \label{linear_sys_4} \\
&c_1+(\tilde J-KQ) d_1-KR d_2-J_{p} K b_1=J_{s} \tilde b_1 \\
&c_2+(\tilde J-K Q) d_2-KR d_1-J_{p} K b_2=J_{s}\tilde b_2 \\
&d_1-(\tilde J-KQ)c_1+K R c_2+J_{p}K a_1=-J_{s} \tilde a_1 \\
&d_2-(\tilde J-KQ)c_2+KR c_1+J_{p}K a_2=-J_{s} \tilde a_2 \,.
\end{align}
The first four equations allow us to express $a_1$, $a_2$, $b_1$, and $b_2$ as linear combinations of $c_1$, $c_2$, $d_1$, and $d_2$. If we plug the resulting expressions into the last four equations we obtain
\begin{equation}
M \left( \begin{array}{c} c_1 \\ c_2 \\ d_1 \\ d_2 \end{array} \right)= J_{s} \left( \begin{array}{c}  \tilde b_1\\ \tilde b_2 \\ -\tilde a_1 \\ -\tilde a_2  \end{array} \right) \,,
\end{equation}
where
\begin{equation}
M=
\begin{pmatrix}
1+J_{p} K Q& \ J_{p}KR& \ \tilde J-KQ& \ -KR \\
J_{p}KR& \ 1+J_{p}KQ& \ -KR& \ \tilde J-KQ \\
KQ-\tilde J& \ KR& \ 1+J_{p}KQ& \ J_{p}KR \\
KR& \ KQ-\tilde J& \ J_{p}KR& \ 1+J_{p}KQ \\
\end{pmatrix} \,.
\end{equation}
So now our problem is to check if $M$ is invertible and to compute $M^{-1}$. Let us set
\begin{align}
M_1&=1+J_{p} K Q \\
M_2&=J_{p}KR \\
M_3&=\tilde J-KQ \\
M_4&=KR \,.
\end{align}
Straightforward computations show that
\begin{equation}
\text{det}(M)=(M_1^2-M_2^2)^2+(M_3^2-M_4^2)^2+2(M_1M_3+M_2M_4)^2+2(M_1 M_4+M_2 M_3)^2 \,.
\end{equation}
We observe that when $\alpha$ is sufficiently small $|Q|>R$ and so $M_3 > M_4$. It follows that $(M_3^2-M_4^2)^2>0$, thus $\text{det}(M)$ is strictly positive. So we can compute the inverse of $M$, and the result is
\begin{equation}
M^{-1}=\frac{1}{\det M}
\begin{pmatrix}
m_1& \ m_2& \ m_3& \ -m_4 \\
m_2& \ m_1& \ -m_4& \ m_3 \\
-m_3& \ m_4& \ m_1& m_2 \\
m_4& \ -m_3& \ m_2& \ m_1
\end{pmatrix} \,,
\end{equation}
where
\begin{align}
m_1&:=M_1^3+2M_2 M_3 M_4+M_1 M_3^2-M_1 M_2^2+M_1 M_4^2 \\
m_2&:=M_2^3+2M_1 M_3 M_4+M_2 M_4^2+M_2M_3^2-M_1^2M_2 \\
m_3&:=-M_3^3-2M_1M_2M_4+M_3M_4^2-M_1^2M_3-M_2^2M_3 \\
m_4&:=-M_4^3-2M_1M_2M_3-M_1^2M_4+M_3^2M_4-M_2^2M_4 \,.
\end{align}
Now we can compute the coefficients $a_i$, $b_i$, $c_i$, $d_i$, $i=1,2$. First we observe that
\begin{equation} \label{c,d}
\left( \begin{array}{c} c_1 \\ c_2 \\ d_1 \\ d_2 \end{array} \right)=J_{s} M^{-1} \left( \begin{array}{c}  \tilde b_1\\ \tilde b_2 \\ -\tilde a_1 \\ -\tilde a_2  \end{array} \right) \,.
\end{equation}
The remaining coefficients can be computed using \eqref{linear_sys_1}-\eqref{linear_sys_4}. At steady state the shape of the system evolves along a closed loop in the configuration space. Figure \ref{loops_comparison} shows a comparison between the loops obtained through numerical simulations and the ones corresponding to the leading order approximation.

\subsection{Numerical simulations}

In this section we present the results of some numerical simulations. We used MATLAB ode45 procedure, which consists of a Runge-Kutta integration scheme with adaptive step size. The values of the dimensionless parameters used in the simulations are shown in Table \ref{parameters}.
\begin{table}[h!] 
\centering
\begin{tabular}{|lllllllll|} 
$J_{s}$  &  $J_{p}$  & $K$ & $\alpha$ & $\tilde a_1$ & $\tilde b_1$ & $\tilde a_2$ & $\tilde b_2$ & $\epsilon$\\ 
4  & 3  & 2 & 0.1 &1 & 0 & 0.25 & 0 & 0.7 \\  
\end{tabular}
\caption{Values of the parameters used in the numerical simulations. \label{parameters}}
\end{table}

\begin{figure}[!h]
\includegraphics[width=1 \textwidth]{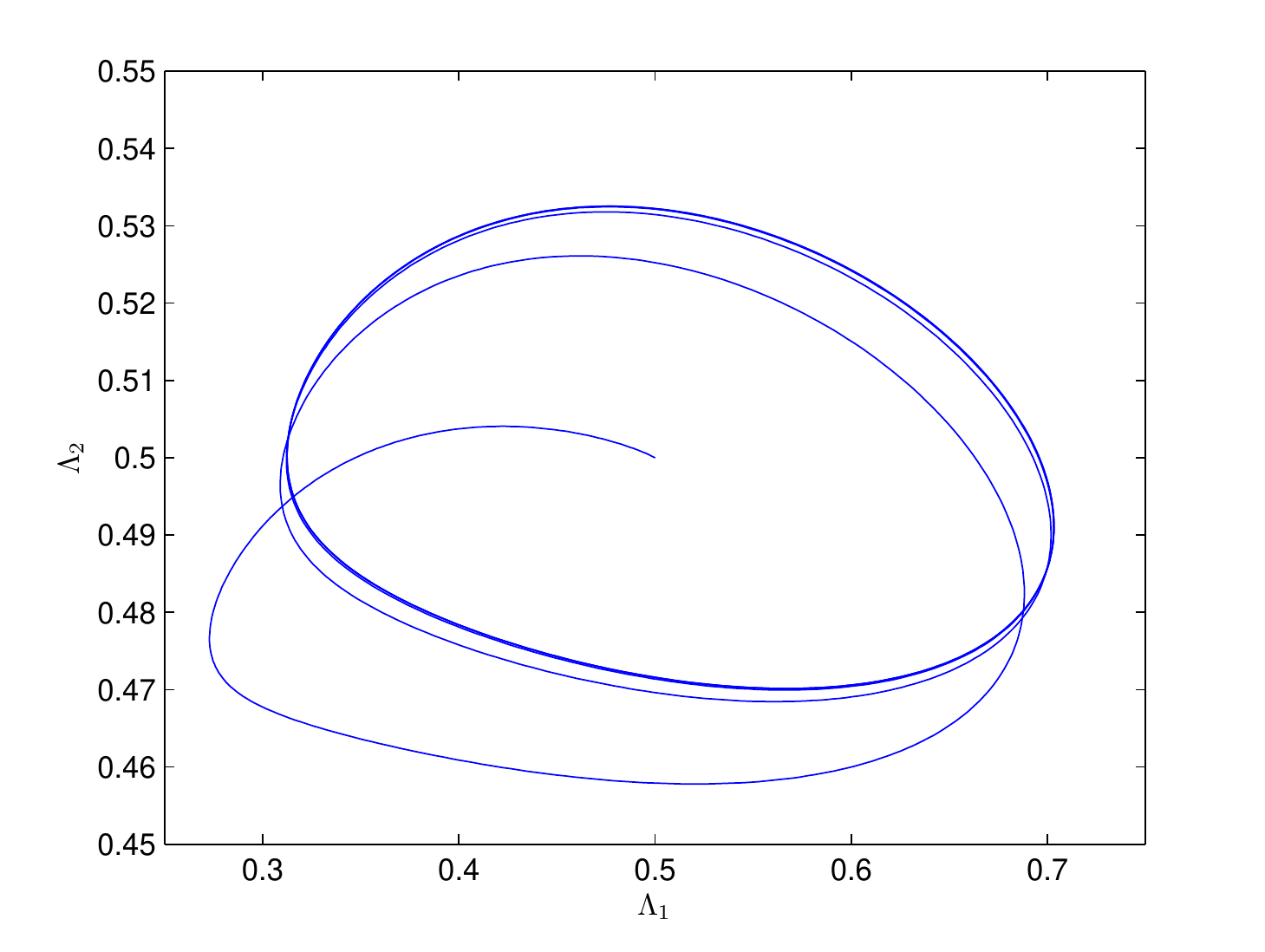}
\caption{TSS with two mucle-like arms: evolution of the two shape parameters $\Lambda_1$ and $\Lambda_2$. We see that the system converges to a closed loop, in agreement with the results of section \ref{qualitative}.}
\label{loop}
\end{figure}

\begin{figure}[!h]
\includegraphics[width=1 \textwidth]{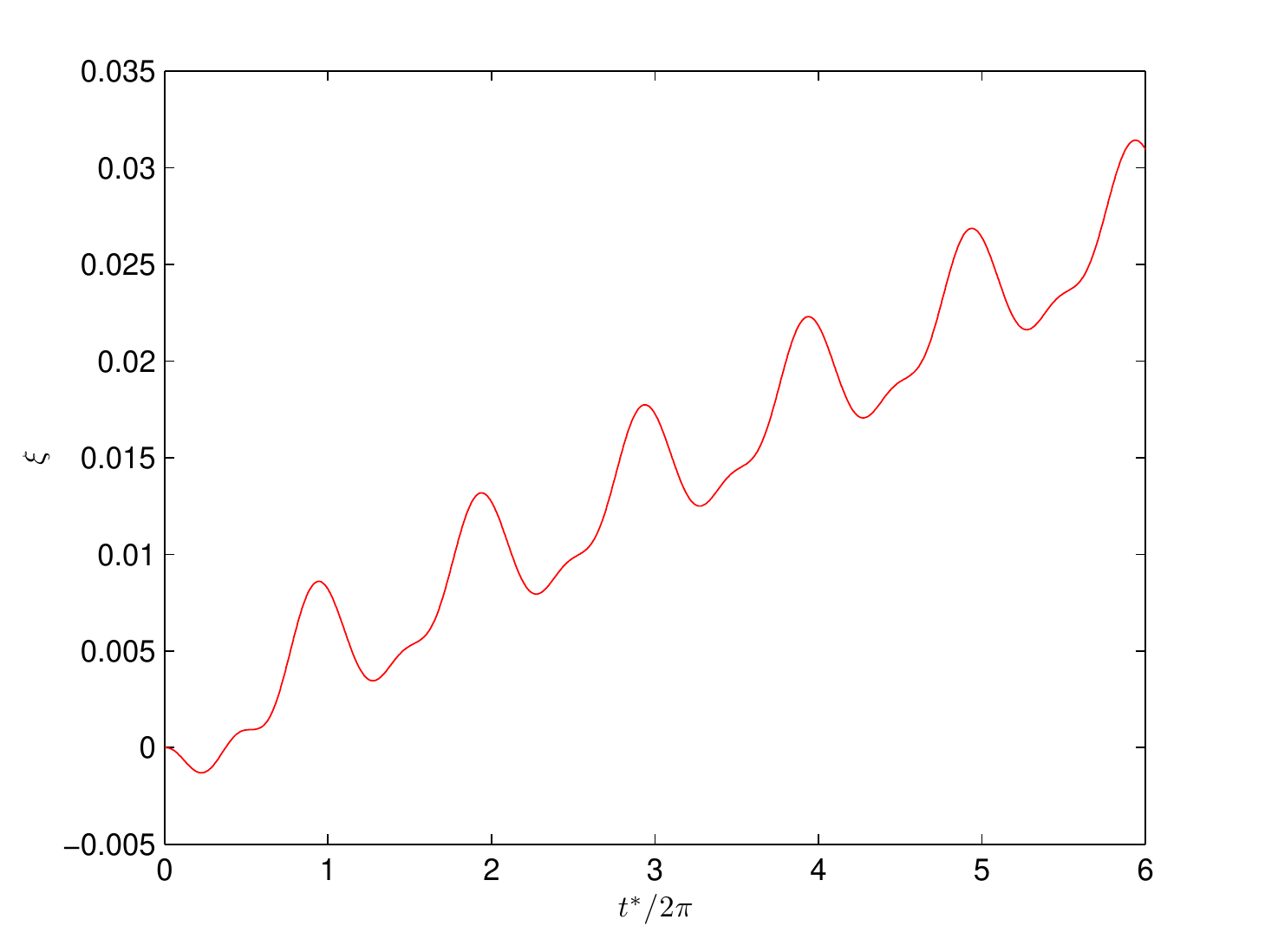}
\caption{TSS with two muscle-like arms: plot of $\xi$ as a function of the normalized time $t^*/2 \pi$. We observe that in this case the net displacement in one period is negative.}
\label{displacement}
\end{figure}

\begin{figure}[!h]
\includegraphics[width=1 \textwidth]{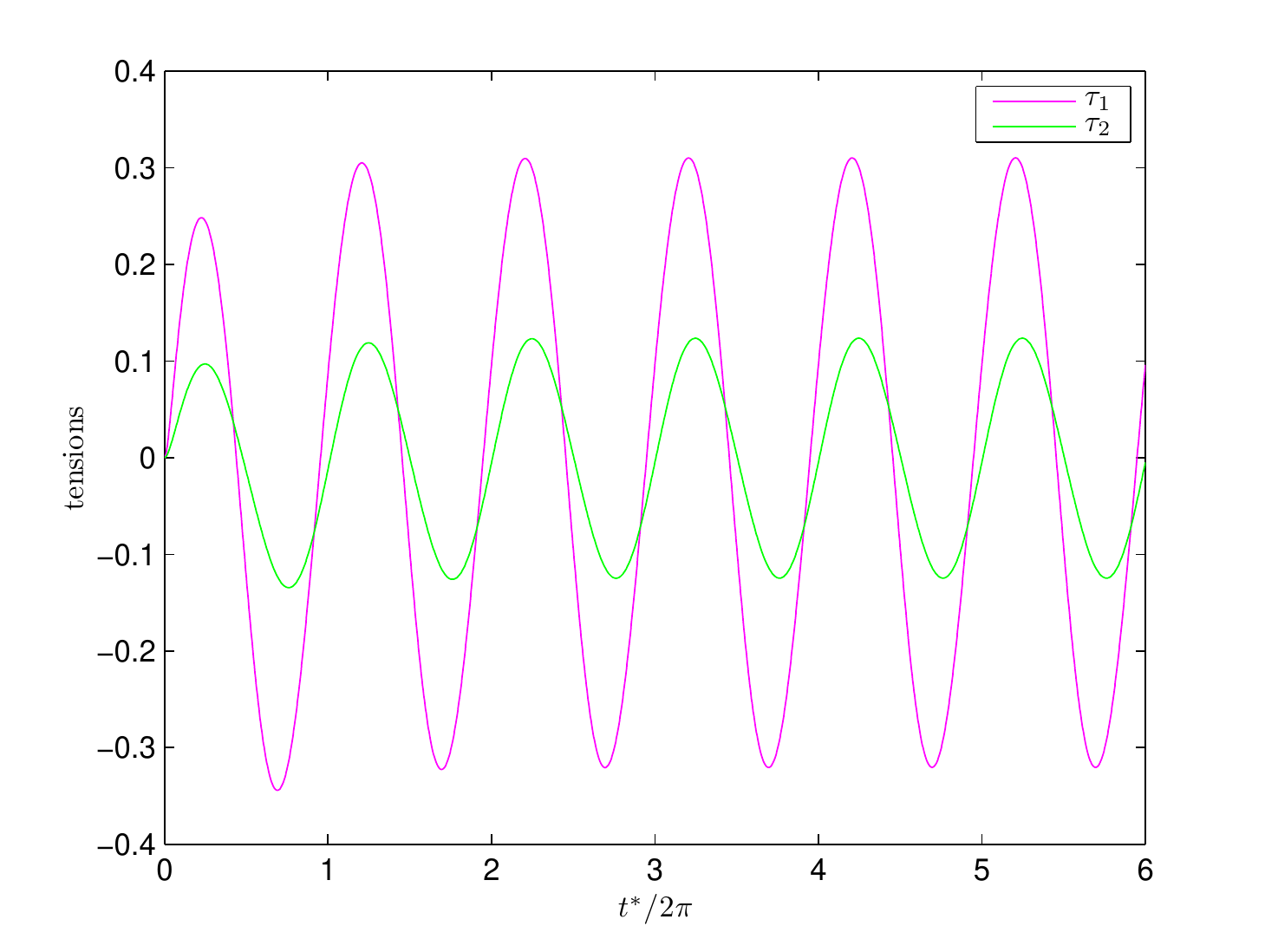}
\caption{TSS with two muscle-like arms: plot of $\tau_1$ and $\tau_2$ as functions of the normalized time $t^*/2 \pi$. The behaviour is periodic, in agreement with the analysis of section \ref{qualitative}.}
\label{tensions}
\end{figure}

In section \ref{qualitative} we proved that the system converges to a periodic orbit. This is confirmed by Figure \ref{loop} in which we see that, after relaxation, $\Lambda_1$ and $\Lambda_2$ evolve along a closed loop. Figure \ref{displacement} shows the evolution of $\xi$. In this simulation $(\tilde a_1, \tilde b_1)$ is proportional to $(\tilde a_2, \tilde b_2)$ (see Table \ref{parameters}). This means that the tensions developed in the active components of the two arms are in phase with each other. If this sinchronization of the active components caused a sinchronization in the length change of the two arms, we would obtain a reciprocal shape change, thus no net motion in view of the scallop theorem. As we can see from figures \ref{loop} and \ref{displacement}, this is not the case. This property is very interesting because it implies that the swimmer can move even if the two active components are stimulated at the same frequency. We will go back to this in the following section, where this property of the system will emerge from the formula for the leading order term of the net displacement in one period.  In Figure \ref{tensions} we see the evolution of $\tau_1$ and $\tau_2$: also in this case the behaviour at steady state is periodic, in agreement with the results of subsection \ref{qualitative}.

\subsection{Optimization} \label{optimization}

In this subsection we study some optimization problems. We consider three performance measures: net displacement in one period, work per travelled distance, and Lighthill's efficiency. The value of the performance measures is determined by the dimensionless parameters that appear in our system of ODEs. These parameters depend on the swimmer's body length $l$, on the actuation frequency $\omega$, on the fluid viscosity $\mu$, and on the physical constants $k_s,k_p,h,B$ characterizing the viscoelastic arms. We consider the case in which the fluid and the swimmer are given, so that the only parameter which can be varied is the actuation frequency $\omega$. We would like to study optimization of the different performance measures with respect to $\omega$. Our strategy is to compute the leading order approximation of the performance measures and use the resulting expressions to study our optimization problem. Then we will compare the optimality results obtained through the leading order approximation with the outcome of numerical simulations.

The first performance measure we consider is $\Delta x$, the net displacement per period in units of body length. This quantity can be computed using formula \eqref{deltax}. Through a simple change of integration variable we see that
\begin{equation} \label{deltax_nondim}
\Delta x=\frac{\alpha}{6} \Big( \frac{1}{\lambda_1^2}+\frac{1}{\lambda_2^2}-1 \Big) \int_0^{\omega T} (u_1 u_2'-u_1' u_2) dt^* \,.
\end{equation}
By using the leading order expressions computed above we find the following leading order approximation for $\Delta x$
\begin{equation} \label{deltax_ab}
\Delta x=\frac{7 \pi \alpha}{3} (a_2 b_1-a_1b_2) \epsilon^2 +O(\epsilon^3) \,.
\end{equation}
By using equations \eqref{linear_sys_1}-\eqref{linear_sys_4} we obtain
\begin{equation}
\Delta x=\frac{7 \pi \alpha}{3} (Q^2-R^2) (c_2 d_1-c_1 d_2) \epsilon^2 +O(\epsilon^3) \,.
\end{equation}
Now by using the expressions for $c_1,c_2,d_1$ and $d_2$ computed in subsection \ref{asymptotic} we obtain
\begin{equation} \label{deltax_final_formula}
\frac{7 \pi \alpha}{3}(Q^2-R^2) \Big( \frac{J_{s}}{\det M} \Big)^2 \Big[(m_1m_4+m_2m_3)(\tilde a_2^2+\tilde b_2^2-\tilde a_1^2-\tilde b_1^2)+(m_1^2-m_2^2+m_3^2-m_4^2)(\tilde b_1 \tilde a_2-\tilde b_2 \tilde a_1)\Big] \epsilon^2 +O(\epsilon^3) \,.
\end{equation}
There are some interesting facts to point out about this formula. First of all we notice that if we change $\sigma_1$ with $\sigma_2$ and viceversa the displacement changes sign, as expected for symmetry reasons. Secondly, we observe that even in the case in which only one of the two arms is activated we obtain a non-zero displacement. Finally, let us comment on the case in which the two active components are sinchronized. Suppose there exists a constant $\eta>0$ such that
\begin{equation}
 \left( \begin{array}{c} \tilde a_2 \\ \tilde b_2 \end{array} \right)= \eta \left( \begin{array}{c}\tilde a_1 \\ \tilde b_1 \end{array} \right) \,.
\end{equation}
In this case one might expect a sinchronization of the length change of the two arms, leading to a reciprocal shape change which produces no net motion. However, our asymptotic analysis shows that is not the case provided that $\eta \ne 1$: from formula \eqref{deltax_final_formula} we obtain that the leading order term of $\Delta x$ is
\begin{equation}
\epsilon^2 \frac{7 \pi \alpha}{3}(Q^2-R^2) \Big( \frac{J_{s}}{\det M} \Big)^2 (m_1m_4+m_2m_3)(\tilde a_1^2+\tilde b_1^2)(\eta-1)  \ne 0 \,.
\end{equation}

Now we compute the mechanical work $\mu \omega l^3 W$ done by the active components in one period. The power expenditure $\mu \omega^2 l^3 \mathcal P$ is
\begin{equation} \label{power}
\mu \omega^2 l^3 \mathcal P =\mu \omega^2 l^3 (\mathcal P_1+\mathcal P_2 ) =S_1 \Big(\frac{\dot T_1}{k_{s,1}}-\dot L_1 \Big)+S_2 \Big(\frac{\dot T_2}{k_{s,2}}-\dot L_2 \Big) \,.
\end{equation} 
It follows that
\begin{equation}
\mu \omega l^3 W= \mu \omega^2 l^3 \int_0^T \mathcal P (t) dt \,.
\end{equation}
Using the leading order approximations computed in subsection \ref{asymptotic} we find that
\begin{equation}
W=\pi \Big[\frac{1}{K}(\tilde b_1 c_1-\tilde a_1 d_1+\tilde b_2 c_2-\tilde a_2 d_2) +\tilde a_1 b_1+\tilde a_2 b_2 -\tilde b_1 a_1-\tilde b_2 a_2\Big] \epsilon^2 +O(\epsilon^3) \,.
\end{equation}

The other two performance measures we consider are are the mechanical work per travelled distance
\begin{equation}
\zeta:=\frac{W}{|\Delta x|}
\end{equation}
and Lighthill's efficiency
\begin{equation}
\eta:=9 \alpha \frac{\Delta x^2}{W} \,.
\end{equation}
Since we have already computed leading order expressions for $\Delta x$ and $W$, we can easily compute leading order expression for $\zeta$ and $\eta$.

Now we would like to optimize the leading order terms of the different performance measures with respect to the actuation frequency $\omega$. First of all we will show that this problem is well-posed, namely, that each performance measure admits an optimal value of $\omega$. Secondly, we will show that the optimality results obtained by studying the leading order approximations are in good agreement with the results of numerical simulations.

\begin{figure}[!h]
\includegraphics[width=1 \textwidth]{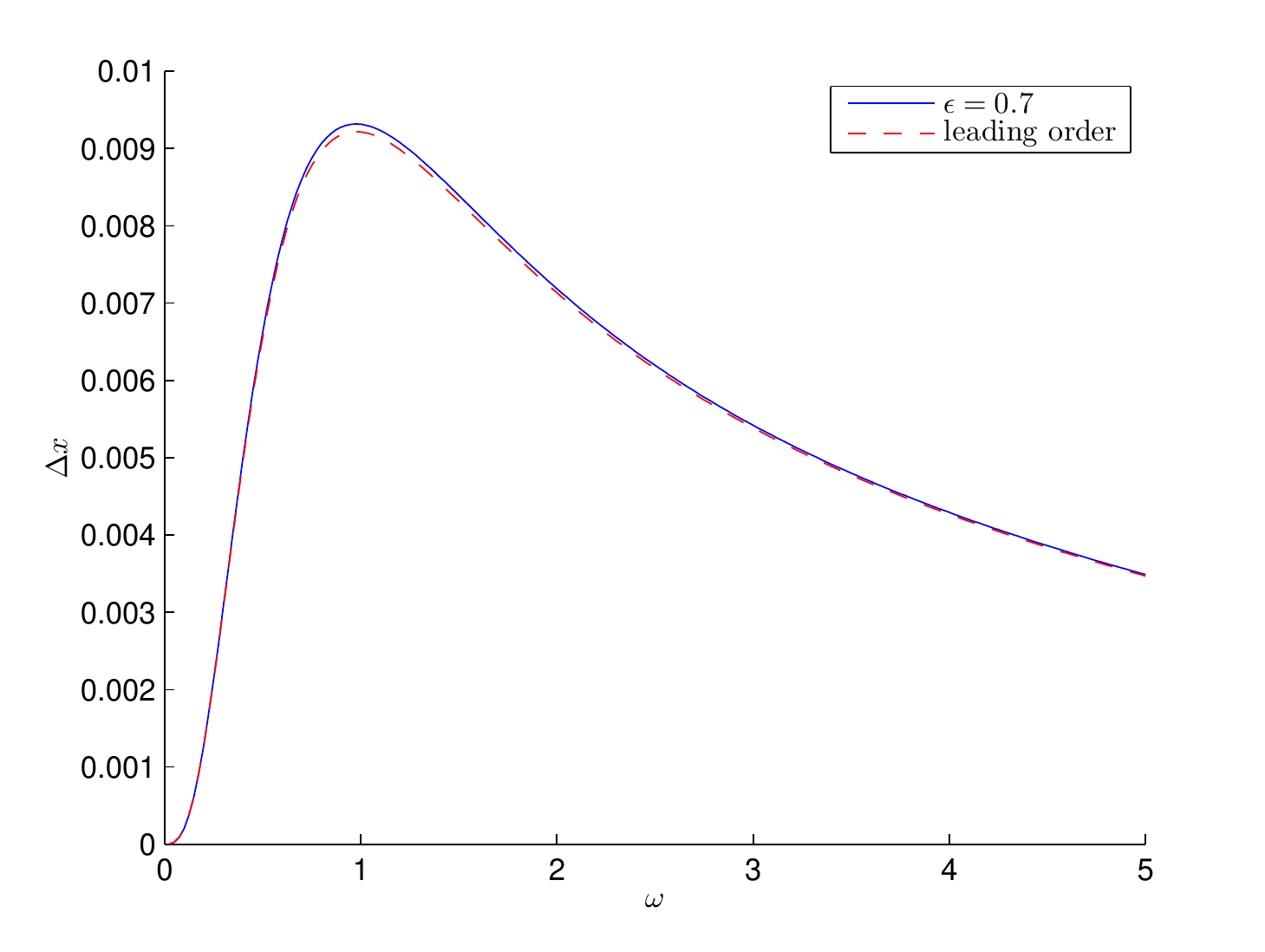}
\caption{Net displacement per period $\Delta x$ as a function of the actuation frequency $\omega$. The blue line corresponds to the results of numerical simulations while the red dashed line corresponds to the leading order approximation. Here the values of the physical constants are such that for $\omega=1$ the dimensionless parameters are as in Table \ref{parameters}.}
\label{optimality1}
\end{figure}

\begin{figure}[!h]
\includegraphics[width=1 \textwidth]{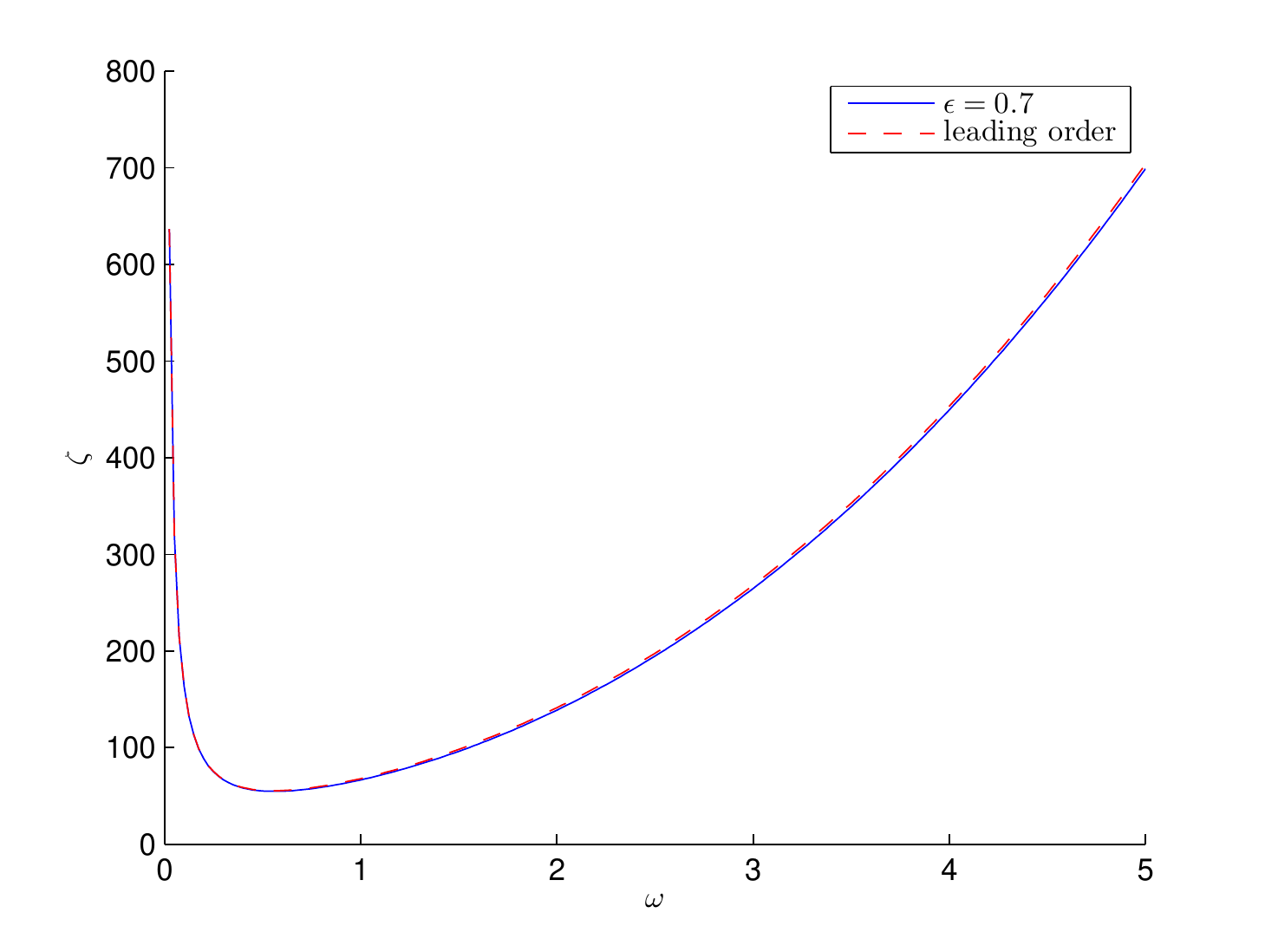}
\caption{Work per travelled distance $\zeta$ as a function of the actuation frequency $\omega$. The blue line corresponds to the results of numerical simulations while the red dashed line corresponds to the leading order approximation. Here the values of the physical constants are such that for $\omega=1$ the dimensionless parameters are as in Table \ref{parameters}.}
\label{optimality2}
\end{figure}

\begin{figure}[!h]
\includegraphics[width=1 \textwidth]{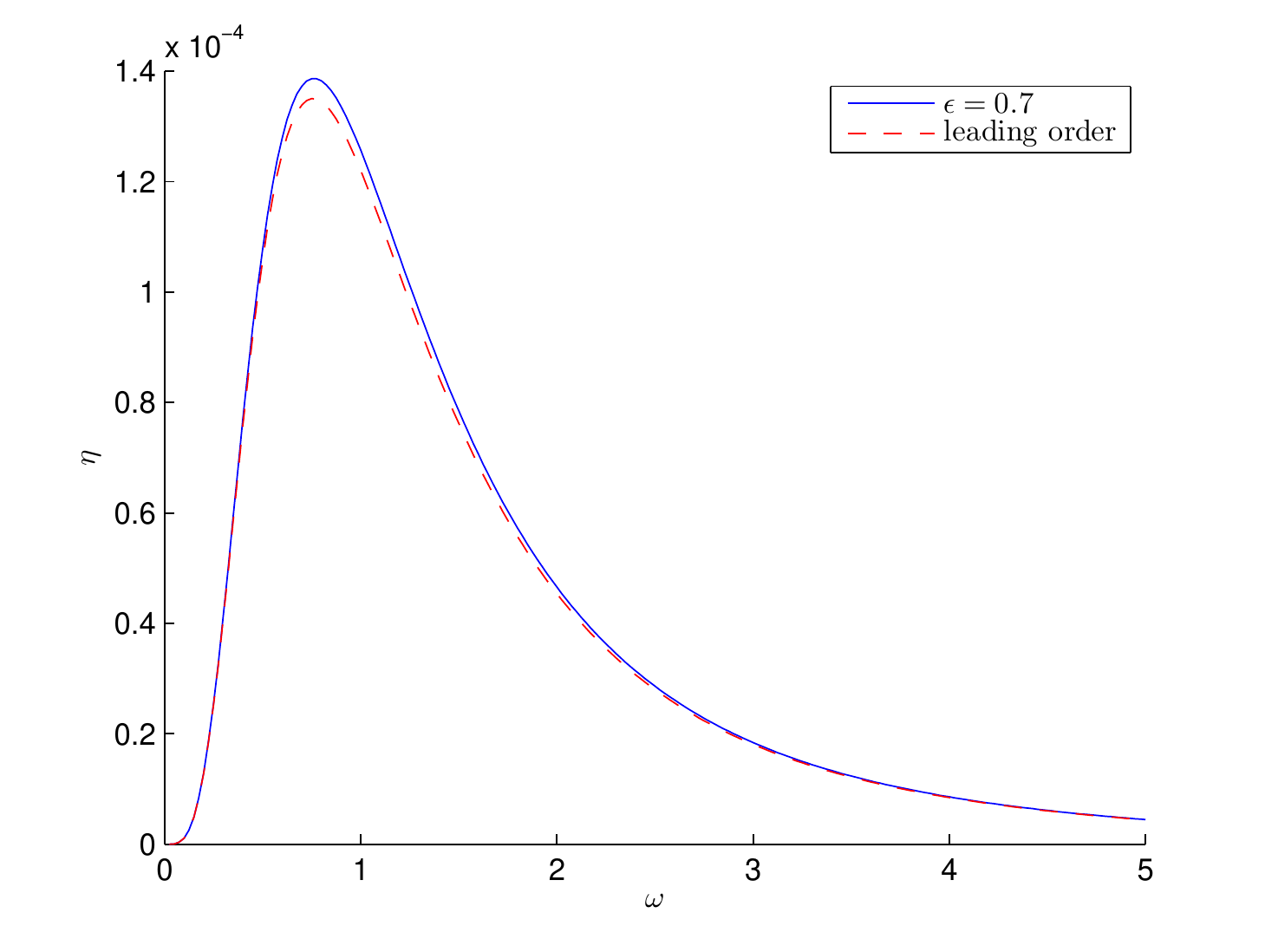}
\caption{Lighthill's efficiency $\eta$ as a function of the actuation frequency $\omega$. The blue line corresponds to the results of numerical simulations while the red dashed line corresponds to the leading order approximation. Here the values of the physical constants are such that for $\omega=1$ the dimensionless parameters are as in Table \ref{parameters}.}
\label{optimality3}
\end{figure}

The coefficients $a_i,b_i,c_i,$ and $d_i$, $i=1,2$, have been computed in subsection \ref{asymptotic}. It is not difficult to check that these coefficients are $O(1/\omega)$ when $\omega \rightarrow + \infty$ and $O(\omega)$ when $\omega \rightarrow 0$. As a consequence, it is easy to verify that 
\begin{align}
\Delta x^{\text{(leading order)}} &=O(1/\omega^2) \text{ for } \omega \rightarrow + \infty \nonumber \\
\Delta x^{\text{(leading order)}} &=O(\omega^2) \text{ for } \omega \rightarrow 0 \nonumber \,.
\end{align}
So the leading order expression for $\Delta x$ vanishes when $\omega$ goes to zero and to $+ \infty$. This implies that there exists $\omega_{\Delta x} \in (0,+\infty)$ which optimizes this performance measure. The situation is similar for the other performance measures:
\begin{align}
\zeta^{\text{(leading order)}}&=O(\omega^2) \text{ for } \omega \rightarrow + \infty \nonumber \\
\zeta^{\text{(leading order)}}&=O(1/\omega) \text{ for } \omega \rightarrow 0 \nonumber \\
\eta^{\text{(leading order)}}&=O(1/\omega^4) \text{ for } \omega \rightarrow + \infty \nonumber \\
\eta^{\text{(leading order)}}&=O(\omega^3) \text{ for } \omega \rightarrow 0 \nonumber \,.
\end{align}
From these asymptotic regimes we deduce the existence of two optimal frequencies $\omega_\zeta,\omega_\eta \in (0,+\infty)$. Figures \ref{optimality1}, \ref{optimality2}, and \ref{optimality3} show plots of the different performance measures as functions of $\omega$. Notice the existence of the optimal frequencies  and the very good agreement between leading order approximations and numerical simulations.

\section{TSS with one muscle-like arm and one passive elastic arm}

In this section we study the TSS with one muscle-like arm and one passive elastic arm. We assume that the active component of the muscle-like arm generates a tension 
\begin{equation}
\sigma(t^*)=\epsilon \tilde \sigma(t^*) \,,
\end{equation}
with $\epsilon<<1$ and $|\tilde \sigma| \le 1$. We assume that $\sigma$ is a periodic functionof time. In the first subsection we study the qualitative properties of the solutions. In the second subsection we compute the the leading order approximation of the solutions in the asymptotic regime $\epsilon<<1$. In the third subsection we present the results of numerical simulations. In the fourth subsection we study some optimization problems.

\subsection{Qualitative properties of the solutions} \label{qualitative_elastic}

For the three-sphere swimmer with one muscle-like arm and one passive elastic arm $\tau_1$ is given by equation \eqref{tau1_elastic}. So we can restrict our attention to the three-dimensional system of ODEs governing the evolution of the two shape variables and $\tau_2$. The state of the system is $Y=(u_1,u_2,\tau_2)^{tr}$ and the problem has the form
\begin{equation}
Y' =f(Y)+\epsilon g(t^*) \,,
\end{equation} 
where $f(Y)$ is given by
\begin{equation}
\left( \begin{array}{c}  \frac{1}{\pi} \Big(\frac{1}{2(\lambda_1+u_1)}-\frac{1}{3 \alpha} \Big) H u_1+\frac{1}{\pi} \Big(\frac{1}{6 \alpha}-\frac{1}{4(\lambda_1+u_1)}-\frac{1}{4(\lambda_2+ u_2)}+\frac{1}{4(\lambda_1+u_1+\lambda_2+u_2)} \Big) \tau_2 \\ \frac{1}{\pi} \Big(\frac{1}{6 \alpha}-\frac{1}{4(\lambda_1+u_1)}-\frac{1}{4(\lambda_2+ u_2)}+\frac{1}{4(\lambda_1+u_1+\lambda_2+u_2)} \Big) H u_1+\frac{1}{\pi}\Big(\frac{1}{2(\lambda_2+u_2)}-\frac{1}{3 \alpha} \Big) \tau_2  \\   J_p K u_2 +[\frac{K}{\pi}\Big(\frac{1}{2(\lambda_2+u_2)}-\frac{1}{3 \alpha} \Big)-\tilde J ] \tau_2+\frac{K}{\pi}  \Big(\frac{1}{6 \alpha}-\frac{1}{4(\lambda_1+u_1)}-\frac{1}{4(\lambda_2+ u_2)}+\frac{1}{4(\lambda_1+u_1+\lambda_2+u_2)} \Big) H u_1 \end{array} \right) \,,
\end{equation}
and
\begin{equation} \label{g_elastic}
g(t^*)= \left( \begin{array}{c} 0 \\ 0 \\ J_p \tilde \sigma(t^*) \end{array} \right) \,.
\end{equation}
The unperturbed system $Y'=f(Y)$ has the equilibrium point $Y=0$. The variational system with respect to this equilibrium point (see \cite{farkas1994periodic}, page 303) is
\begin{equation} \label{variational_system_elastic}
Z'= B Z \,,
\end{equation}
with
\begin{equation} \label{B}
B=
\begin{pmatrix}
PH& \ 0& \ R \\
RH& \ 0& \ Q \\
KHR& \ J_{p}K& \ KQ-\tilde J \\
\end{pmatrix} \,,
\end{equation}
where
\begin{align}
P&=\frac{1}{\pi} \Big(\frac{1}{2 \lambda_1}-\frac{1}{3 \alpha} \Big) \\
Q&= \frac{1}{\pi} \Big(\frac{1}{2 \lambda_2}-\frac{1}{3 \alpha} \Big) \\
R&= \frac{1}{\pi} \Big( \frac{1}{6 \alpha}-\frac{1}{4 \lambda_1}-\frac{1}{4 \lambda_2}+\frac{1}{4} \Big) \,.
\end{align}
The characteristic polynomial of $B$ is
\begin{equation} \label{characteristic_poly_elastic}
P_B(x)=x^3+[\tilde J-KQ-P H] x^2+[K H(PQ-R^2)-P \tilde J H-J_{p}KQ]x+J_{p}KH(PQ-R^2) \,.
\end{equation}
If we prove that all the eigenvalues of $B$ have negative real part then we can argue as in section \ref{qualitative} and conclude that, for $\epsilon$ small enough, there exists one and only one periodic orbit which is asymptotically stable.
Let us call $a_j$, $j=0,1,2$ the coefficients of the characteristic polynomial, so that
\begin{equation}
p_B(x)=x^3+a_2x^2+a_1x+a_0 \,.
\end{equation}
If $\alpha$ is small enough then $P<0$, $Q<0$, $R>0$, $|P|<R$, and $|Q|<R$. So under this hypothesis all the coefficients of the polynomial are positive. The table associated to the polynomial by means of the Routh method is
\begin{equation} \label{table_Routh_2}
\begin{bmatrix}
1 & \ a_1 \\
a_2& \ a_0 \\
b_2& \ ... \\
\end{bmatrix} \,,
\end{equation}
with
\begin{equation}
b_2=(a_1 a_2-a_0)/a_2 \,.
\end{equation}
Notice that
\begin{align}
a_1a_2-a_0 &=(\tilde J-KQ-PH)(KH(PQ-R^2)-P \tilde J H-J_{p}KQ)-J_{p}KH(PQ-R^2)\nonumber \\
&=(J_{s}-QK-PH)(KH(PQ-R^2)-P H \tilde J-QKJ_{p})+...\nonumber \\
&...+J_{p}(-P H \tilde J -QKJ_{p})>0  \,.
\end{align}
It follows that $b_2>0$. So by applying Routh's criterion we see from table \eqref{table_Routh_2} that there are two roots with negative real part. Since the determinant of $B$ is negative, also the third root must have negative real part.

\subsection{Asymptotic expansions} \label{asymptotic_elastic}

\begin{figure}[!h]
\includegraphics[width=1 \textwidth]{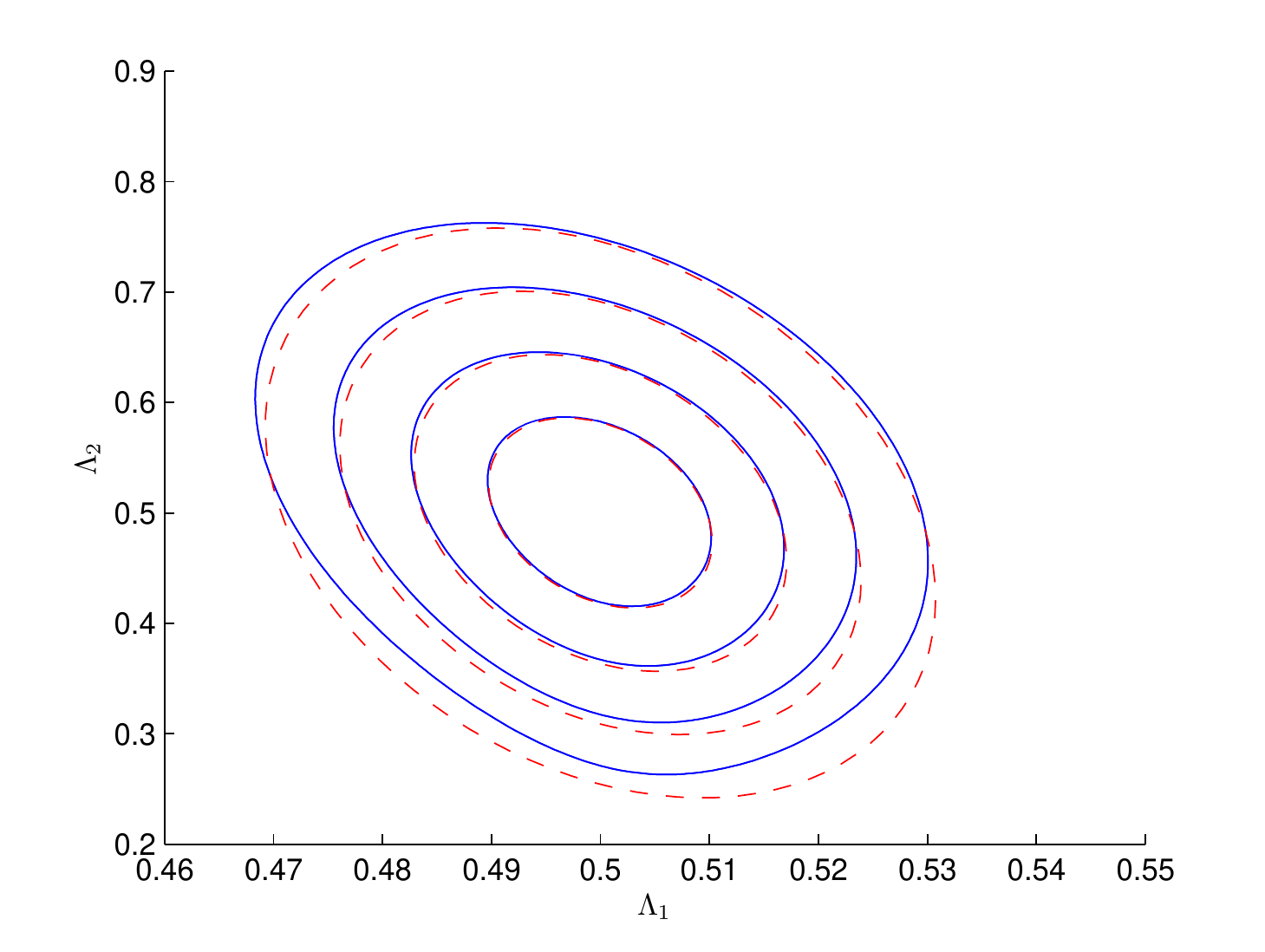}
\caption{TSS with one muscle-like arm and one passive elastic arm: the closed curves are the loops along which the two shape parameters evolve. The full blue lines show the results of numerical simulations while the red dashed lines are the leading order approximations. These loops are obtained by varying $\epsilon$ and choosing the other parameters as in Table \ref{parameters_elastic}. The different loops, from the smallest to the largest, correspond to $\epsilon=0.3$, $\epsilon=0.5$, $\epsilon=0.7$, and $\epsilon=0.9$. We can see from this picture that the error becomes vanishingly small for small values of $\epsilon$.}
\label{loops_comparison_elastic}
\end{figure}

Now we would like to compute analytically the leading order term of the solutions. Let us express the solutions in power series as follows
\begin{align}
u_1&=\epsilon u_1^{(1)}+\epsilon^2 u^{(2)}_1+... \\
u_2&=\epsilon u_2^{(1)}+\epsilon^2 u_2^{(2)}+... \\
\tau_2&=\epsilon \tau_2^{(1)}+\epsilon^2 \tau_2^{(2)}+...
\end{align}
Let us set $X=(u_1^{(1)},u_2^{(1)},\tau_2^{(1)})^{tr}$. $X(t^*)$ is the solution of
\begin{equation}
\begin{cases}
&X'=BX+g(t^*) \\
&X(0)=0 \,.
\end{cases}
\end{equation}
The solution of this problem can be computed with the help of Duhamel's formula. Since we are interested in the periodic behaviour of the solution at steady state, we would like to compute the periodic orbit of the system. We assume that
\begin{equation}
\tilde \sigma (t^*)=\tilde a \sin(t^*)+\tilde b \cos( t^*) \,.
\end{equation}
We can write $g$ as follows
\begin{equation}
g(t^*)=\hat g_s \sin(t^*)+\hat g_c \cos(t^*) \,,
\end{equation}
where $\hat g_s=(0,0,J_s \tilde a)^{tr}$ and $\hat g_c=(0,0,J_s \tilde b)^{tr}$. We look for the periodic solution in the following form
\begin{equation}
X(t^*) =\hat X_s\sin(t^*)+\hat X_c \cos(t^*) \,,
\end{equation}
with $\hat X_s,\hat X_c \in \mathbb{R}^3$. So our differential equation becomes
\begin{equation}
-\hat X_c \sin(t^*)+\hat X_s \cos(t^*)=(B \hat X_s+\hat g_s) \sin(t^*)+(B \hat X_c+\hat g_c) \cos(t^*) \,.
\end{equation}
Since the equality must hold for every time $t^*$, we obtain the following linear system
\begin{equation}
\begin{cases}
&-\hat X_c=B \hat X_s+\hat g_s \\
&\hat X_s=B \hat X_c+ \hat g_c \,.
\end{cases}
\end{equation}
We can rewrite the system in compact form
\begin{equation} \label{linear_system_compact_elastic}
\begin{pmatrix}
-B& \ -\mathbb{1}& \\
\mathbb{1}& \ -B \\
\end{pmatrix} \left( \begin{array}{c} \hat X_s \\ \hat X_c \end{array} \right)=\left( \begin{array}{c} \hat g_s \\ \hat g_c \end{array} \right) \,.
\end{equation}
Now suppose that
\begin{align}
u_1^{(1)}&=a_1 \sin(t^*) +b_1 \cos(t^*) \\
u_2^{(1)}&= a_2 \sin(t^*)+b_2 \cos(t^*) \\
\tau_2^{(1)}&= c \sin(t^*)+d \cos(t^*) \,,
\end{align}
namely, $\hat X_s=(a_1,a_2,c)^{tr}$ and $\hat X_c=(b_1,b_2,d)^{tr}$.
Let us rewrite \eqref{linear_system_compact_elastic} explicitly
\begin{align}
&PHa_1+Rc+b_1=0 \\
&RHa_1+Qc+b_2=0 \\
&a_1-PHb_1-Rd=0 \\
&a_2-RHb_1-Qd=0 \\
&-RKHa_1-KJ_pa_2+(\tilde J-KQ)c-d=J_s \tilde a \\
&c-RKHb_1-KJ_pb_2+(\tilde J-KQ)d=J_s \tilde b \,.
\end{align}
From the first four equations we obtain
\begin{align}
&a_1=-\frac{PRH}{1+P^2H^2}c+\frac{R}{1+P^2H^2} d \label{ale001} \\
&a_2=RH \Big(\frac{P^2RH^2}{1+P^2H^2}-R\Big) c+\Big(Q-\frac{PR^2H^2}{1+P^2H^2}\Big)d \label{ale002} \\
&b_1=\Big(\frac{P^2RH^2}{1+P^2H^2}-R\Big)c-\frac{PRH}{1+P^2H^2} d \label{ale003} \\
&b_2=\Big(\frac{PR^2H^2}{1+P^2H^2}-Q\Big)c-\frac{R^2 H}{1+P^2H^2} d \label{ale004} \,.
\end{align}
If we plug these expressions into the last two equations we obtain the following problem for $c$ and $d$
\begin{equation} \label{linear_problem_N}
N \left( \begin{array}{c} c \\ d  \end{array} \right)= J_s \left( \begin{array}{c}  \tilde a\\ \tilde b  \end{array} \right) \,,
\end{equation}
where
\begin{equation}
N=
\frac{1}{1+P^2H^2} 
\begin{pmatrix}
N_1 & \ -N_2 & \\
N_2 & \ N_1 & \\
\end{pmatrix} \,,
\end{equation}
and
\begin{align}
N_1&=(J_p+J_s)(1+P^2H^2)+KH^2(PR^2-P^2Q)+KHJ_pR^2-KQ \\
N_2&=1+H^2P^2+KHR^2+KH^2J_p(P^2Q-PR^2)+QKJ_p \,.
\end{align}
The linear problem \eqref{linear_system_compact_elastic} has a solution if and only if \eqref{linear_problem_N} has a solution. So we need to check if $N$ is invertible. We notice that
\begin{equation}
\det (N)=\frac{N_1^2+N_2^2}{(1+P^2H^2)^2} >0 \,,
\end{equation}
hence $M$ is invertible. The inverse of $M$ is given by
\begin{equation}
N^{-1}=\frac{1}{\det(N)} \text{cof}(N)^{tr}=\frac{1+P^2H^2}{N_1^2+N_2^2} \begin{pmatrix}
N_1 & \ N_2 & \\
-N_2 & \ N_1 & \\
\end{pmatrix} \,.
\end{equation}
Now $c$ and $d$ can be computed as follows
\begin{equation}
 \left( \begin{array}{c} c \\ d  \end{array} \right)= J_{s} N^{-1} \left( \begin{array}{c}  \tilde a\\ \tilde b  \end{array} \right) \,.
\end{equation}
Once we know $c$ and $d$ we can compute  $a_1,a_2,b_1,b_2$ by using equations \eqref{ale001}-\eqref{ale004}. At steady state the two shape variables evolve along a closed loop in the configuration space. Figure \ref{loops_comparison_elastic} shows a comparison between the closed curves obtained through numerical simulations and the ones corresponding to the leading order approximation.

\subsection{Numerical simulations}

\begin{figure}[!h]
\includegraphics[width=1 \textwidth]{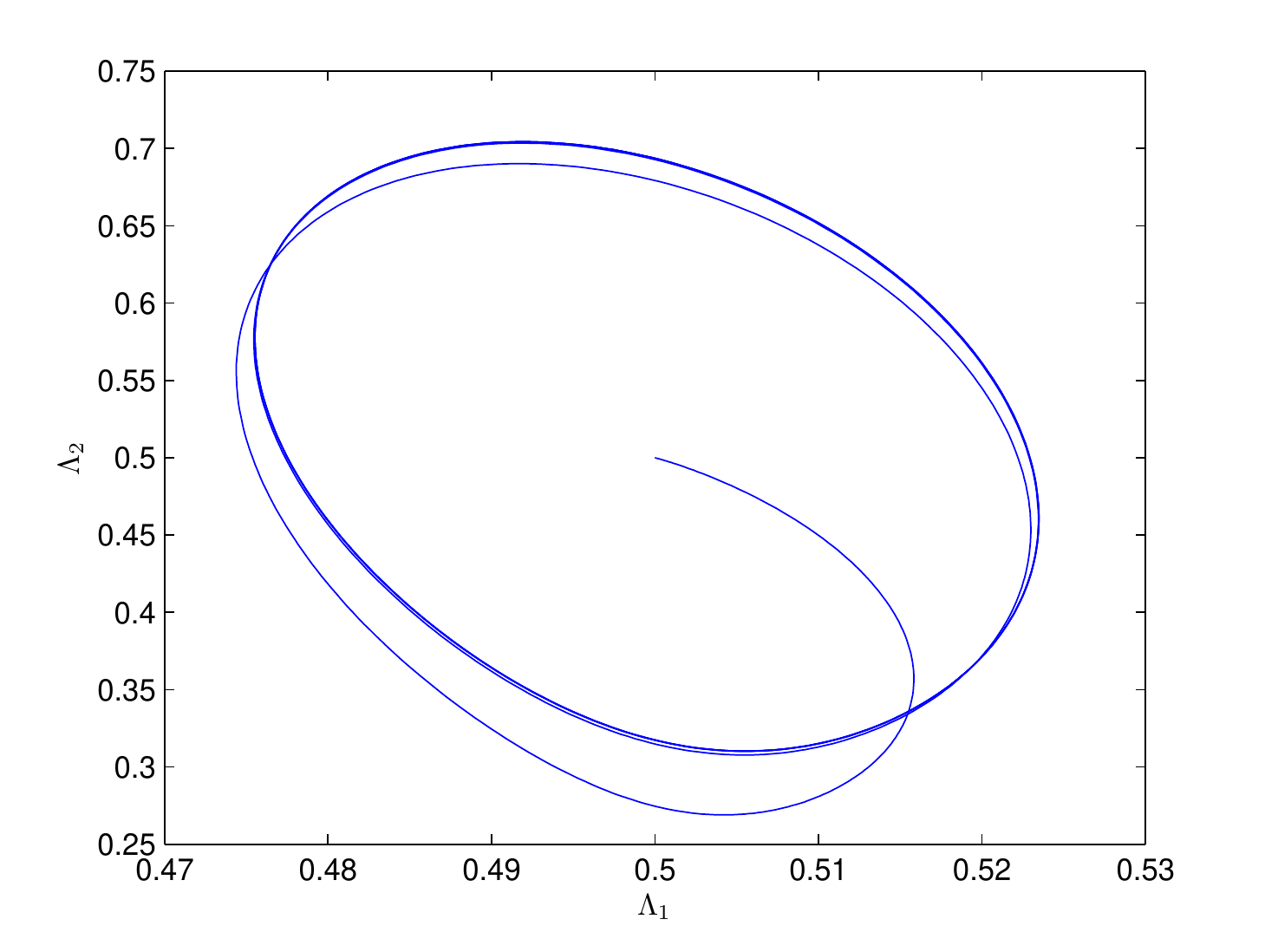}
\caption{TSS with one muscle-like arm and one passive elastic arm: evolution of $\Lambda_1$ and $\Lambda_2$. The solution converges to a closed loop, in agreement with the results of section \ref{qualitative_elastic}.}
\label{loop_elastic}
\end{figure}

\begin{figure}[!h]
\includegraphics[width=1 \textwidth]{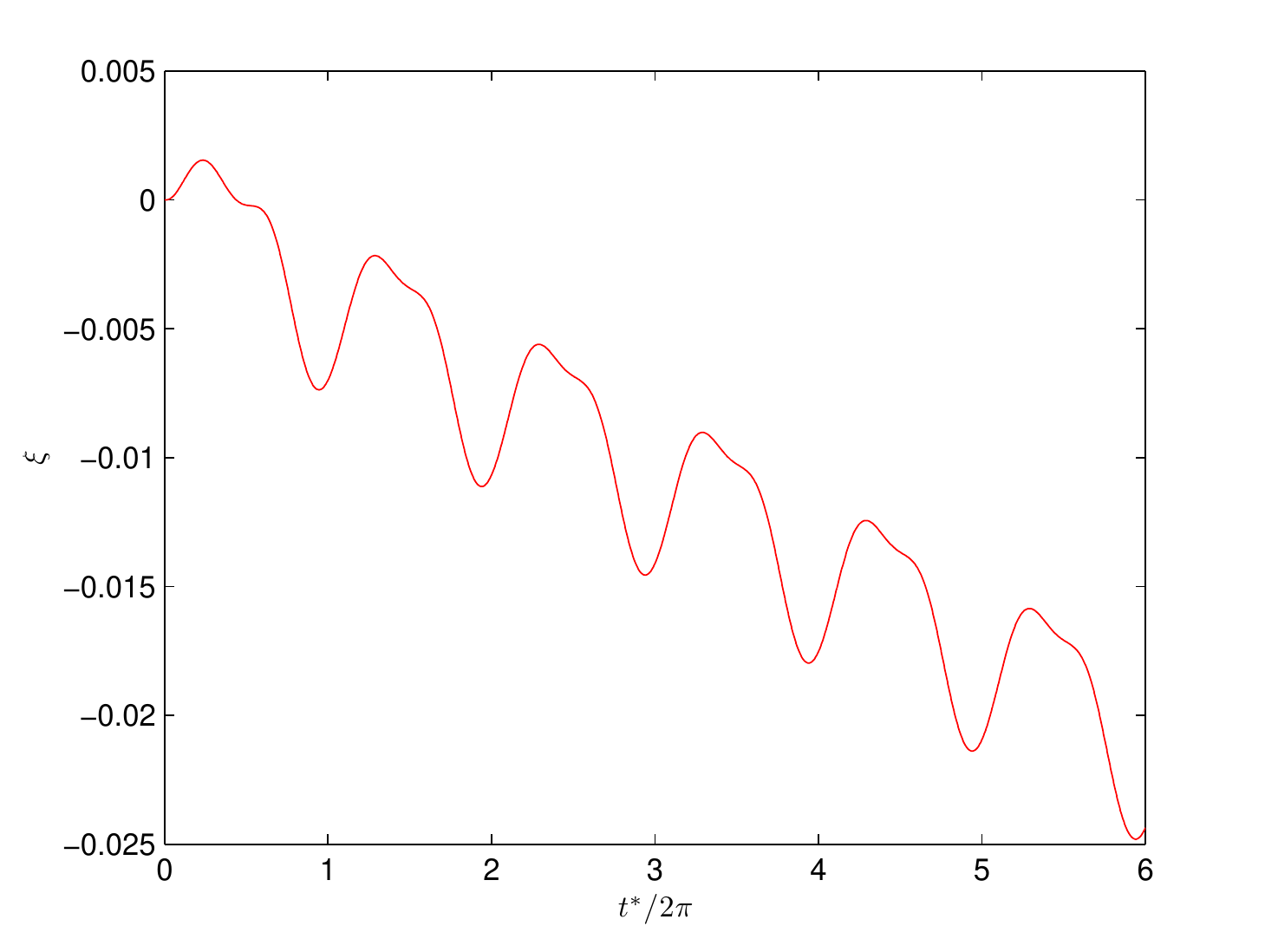}
\caption{TSS with one muscle-like arm and one passive elasic arm: plot of $\xi$ as a function of the normalized time $t^*/2 \pi$. The net displacement in one period is negative.}
\label{displacement_elastic}
\end{figure}

\begin{figure}[!h]
\includegraphics[width=1 \textwidth]{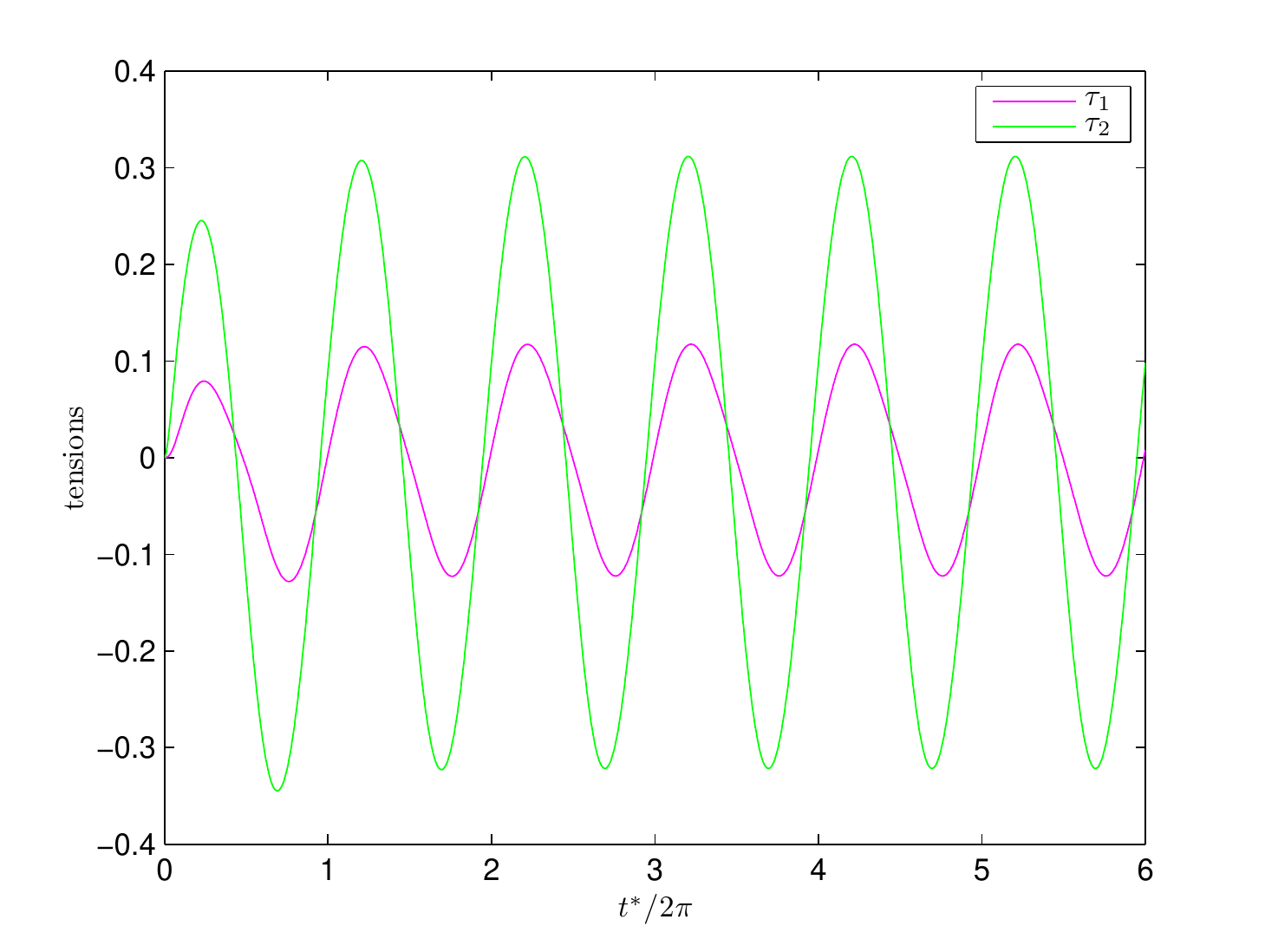}
\caption{TSS with one muscle-like arm and one passive elastic arm: plot of $\tau_1$ and $\tau_2$ as functions of the normalized time $t^*/2 \pi$. The behaviour is periodic, in agreement with the analysis of section \ref{qualitative_elastic}.}
\label{tensions_elastic}
\end{figure}

In this section we present the results of some numerical simulations, obtained by choosing the dimensionless parameters as in Table \ref{parameters_elastic}.
\begin{table}[h!] 
\centering
\begin{tabular}{|llllllll|}
$J_{s}$  & $J_p$ & $K$ & $H$ & $\alpha$ & $\tilde a$ & $\tilde b$ & $\epsilon$ \\
4 & 3 & 2 & 5& 0.1 & 1 & 0 & 0.7
\end{tabular}
\caption{Values of the parameters used in the numerical simulations. \label{parameters_elastic}}
\end{table}
Figure \ref{loop_elastic} shows the evolution of the system in the shape space. We observe that, after relaxation, $\Lambda_1$ and $\Lambda_2$ evolve along a closed loop. In  Figure \ref{displacement_elastic} we see the evolution of $\xi$. We notice that the net displacement in one period has a negative sign, which means that the object swims with the passive arm ahead. This is a general property of the TSS with passive elastic tail. Given any periodic deformation of the active arm, it can be shown that the corresponding net displacement in one period has a negative sign \cite{MDS2015}. Figure \ref{tensions_elastic} shows a plot of $\tau_1$ and $\tau_2$ as functions of time. The behaviour at steady state is periodic, as expected from the results of subsection \ref{qualitative_elastic}.

\subsection{Optimization}

In this subsection we study some optimization problems for the three-sphere swimmer with one muscle-like arm and one passive elastic arm. The optimality measures we consider are the net displacement in one period $\Delta x$, the mechanical work per travelled distance $\zeta$ and Lighthill's efficiency $\eta$. Like in section \ref{optimization}, we assume that the fluid and swimmer are given, so that the only parameter which can be varied is the actuation frequency $\omega$. First we study our optimization problem using the leading order approximation of the different performance measures and then we compare the results with the outcome of numerical simulations.

We start by computing the leading order expressions for the different performance measures. Formula \eqref{deltax_ab} for the leading order approximation of $\Delta x$ remains valid. Let us consider the mechanical work $\mu \omega l^3 W_2$ done by the active component of the muscle-like arm in one period. The power expenditure is
\begin{equation} \label{power_elastic}
\mu \omega^2 l^3 \mathcal P _2= S_2 \Big(\frac{\dot T_2}{k_{s,2}}-\dot L_2 \Big) \,.
\end{equation}
It follows that
\begin{equation}
\mu \omega l^3 W_2= \mu \omega^2 l^3 \int_0^T \mathcal P_2(t) dt \,.
\end{equation}
So by using the results of section \ref{asymptotic_elastic} we obtain
\begin{equation}
W=\pi \Big[ \frac{1}{K} (\tilde b c_2-\tilde a d_2) + \tilde a b_2-\tilde b a_2 \Big] \epsilon^2 +O(\epsilon^3) \,.
\end{equation}
The work per travelled distance $\eta$ and Lighthill's efficiency $\eta$ are defined as in section \ref{optimization}, and the corresponding leading order approximations can be easily obtained once we know the approximation of $\Delta x$ and $W$.

\begin{figure}[!h]
\includegraphics[width=1 \textwidth]{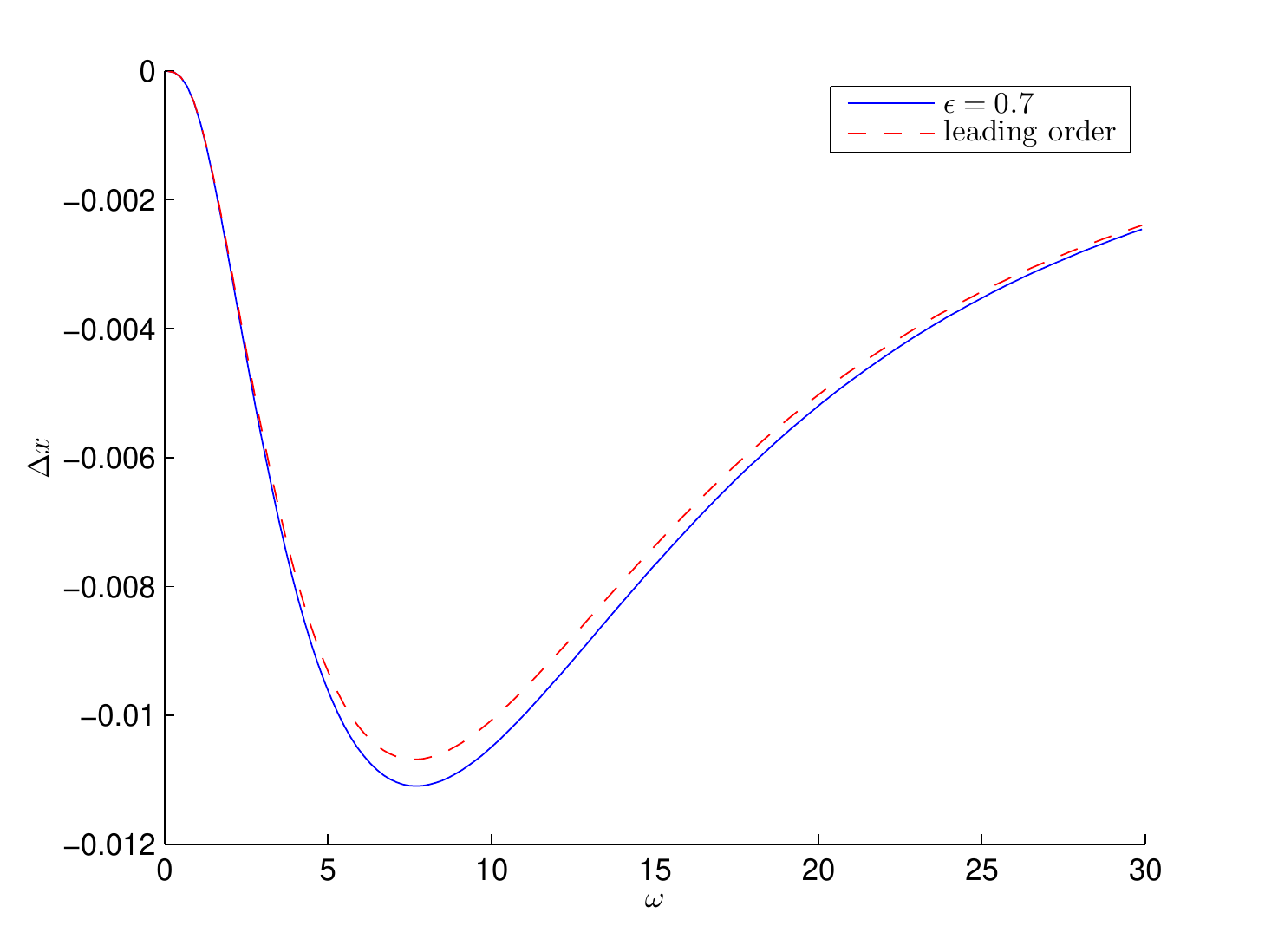}
\caption{Net displacement per period $\Delta x$ as a function of the actuation frequency $\omega$ for the TSS with one muslce-like arm and one passive elastic arm. The blue line corresponds to the results of numerical simulations while the red dashed line corresponds to the leading order approximation. Here the values of the physical constants are such that for $\omega=1$ the dimensionless parameters are as in Table \ref{parameters_elastic}.}
\label{optimality1_elastic}
\end{figure}

\begin{figure}[!h]
\includegraphics[width=1 \textwidth]{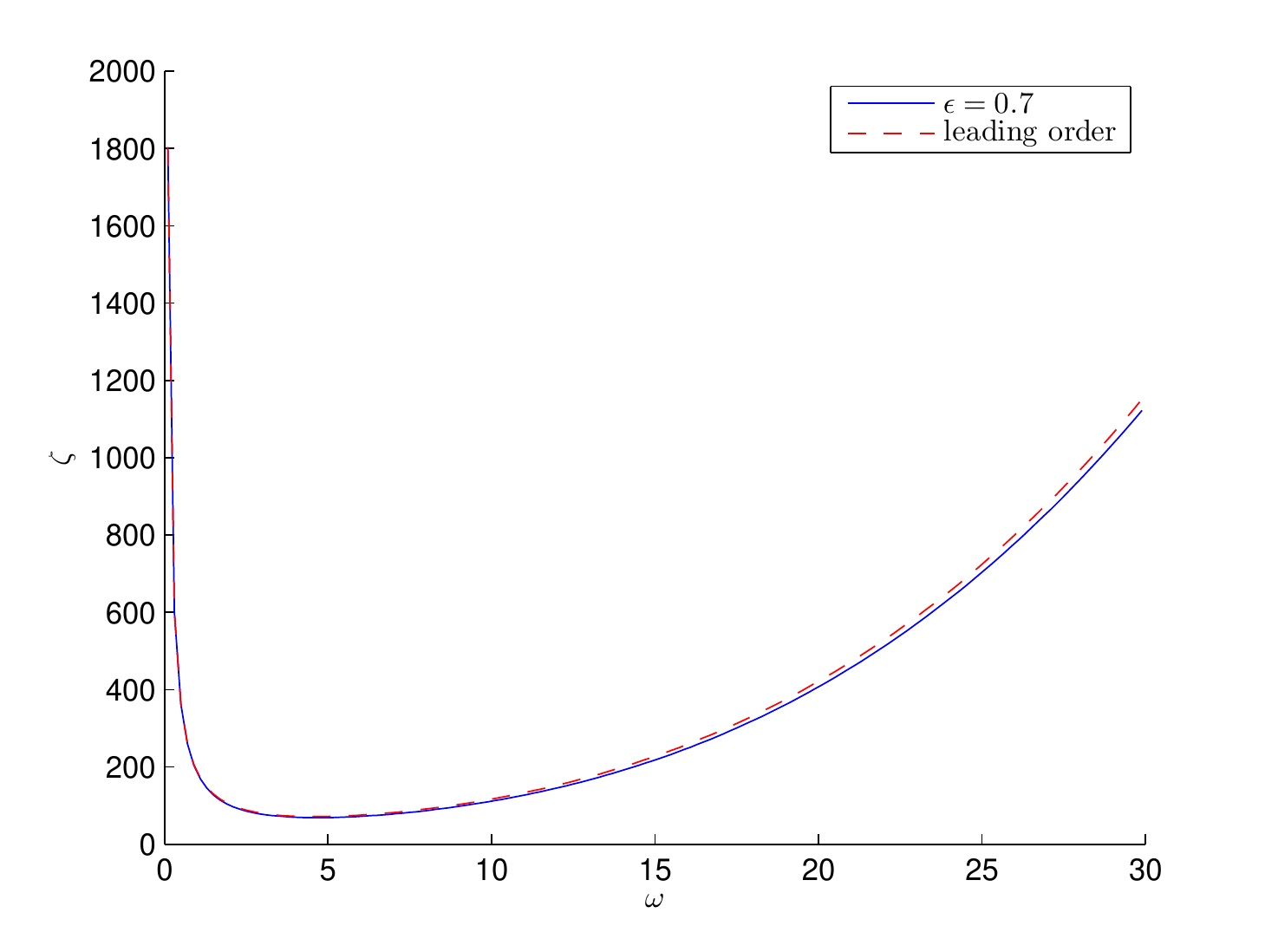}
\caption{Work per travelled distance $\zeta$ as a function of the actuation frequency $\omega$ for the TSS with one muslce-like arm and one passive elastic arm. The blue line corresponds to the results of numerical simulations while the red dashed line corresponds to the leading order approximation. Here the values of the physical constants are such that for $\omega=1$ the dimensionless parameters are as in Table \ref{parameters_elastic}.}
\label{optimality2_elastic}
\end{figure}

\begin{figure}[!h]
\includegraphics[width=1 \textwidth]{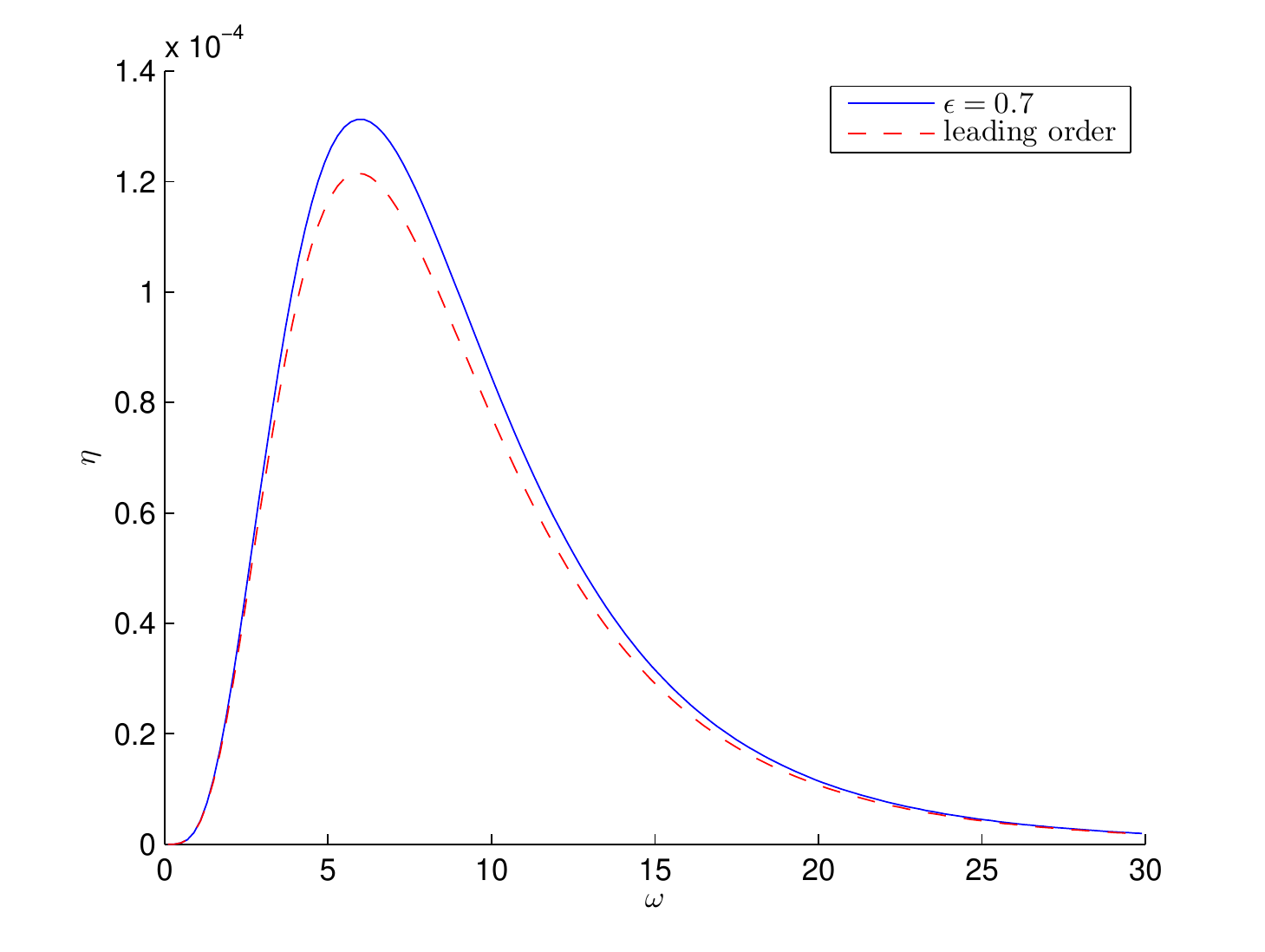}
\caption{Lighthill's efficiency $\eta$ as a function of the actuation frequency $\omega$ for the TSS with one muslce-like arm and one passive elastic arm. The blue line corresponds to the results of numerical simulations while the red dashed line corresponds to the leading order approximation. Here the values of the physical constants are such that for $\omega=1$ the dimensionless parameters are as in Table \ref{parameters_elastic}.}
\label{optimality3_elastic}
\end{figure}

Using the formulas computed in section \ref{asymptotic_elastic} it is not difficult to see that the coefficients $c,d,a_i,b_i$, $i=1,2$ are $O(1/ \omega)$ when $\omega \rightarrow + \infty$ and $O(\omega)$ when $\omega \rightarrow 0$. Therefore, the asymptotic behaviour of the leading order terms of the different performance measures is exactly the same as the one observed in section \ref{optimization}. This implies existence of three optimal frequencies $\omega_{\Delta x}$, $\omega_\zeta$, and $\omega_\eta$. These theoretical predictions are confirmed by Figures \ref{optimality1_elastic}, \ref{optimality2_elastic}, and \ref{optimality3_elastic}. In these figures we observe the existence of the three optimal frequencies $\omega_{\Delta x}$, $\omega_\zeta$, $\omega_\eta$ and the high accuracy of the leading order approximations.

\section*{Conclusions}

We studied the dynamics of the three-sphere swimmer with muscle-like arms. We assumed that the forces generated by the swimmer in the active components of the arms have intensity $\epsilon$ and vary periodically with frequency $\omega$. We showed that the two shape parameters and the forces acting across the arms evolve according to a system of ODEs. We proved that the solutions converge to a periodic orbit. Under the assumption that $\epsilon <<1$, we computed the leading order approximation of the solutions at steady state. Then we studied some optimization problems. We considered three different performance measures: net displacement in one period, work per travelled distance, and Lighthill's efficiency. We studied optimization of these quantities with respect to the actuation frequency $\omega$. We computed leading order approximations of the different performance measures. Using these approximations we showed that each performance measure admits an optimal frequency. In addition we showed that the optimality results obtained through leading order approximations are in very good agreement with the outcome of numerical simulations. Then we introduced the three-sphere swimmer with one muscle-like arm and one passive elastic arm. We studied this model through the same type of analysis done for the three-sphere swimmer with two muscle-like arms.

\appendix

\section{Asymptotic expansions when the forces in the active components are generic periodic functions}

\subsection{TSS with two muscle-like arms}

In section \ref{asymptotic} we computed the leading order approximation of the solution in the case in which $\tilde \sigma_1$ and $\tilde \sigma_2$ are given by equations \eqref{sigmatilde1} and \eqref{sigmatilde2}. Now we would like to consider the case in which $\tilde \sigma_1$ and $\tilde \sigma_2$ are two generic period functions with period $2 \pi$. We consider the expansion in Fourier series
\begin{align}
\tilde \sigma_1 (t^*) &= \sum_{j=1}^{+\infty} \tilde a_{1,j} \sin(j t^*)+\tilde b_{1,j} \cos(j t^*) \\
\tilde \sigma_2(t^*)&=\sum_{j=1}^{+\infty} \tilde a_{2,j} \sin(j t^*)+\tilde b_{2,j} \cos(j t^*) \,.
\end{align}
Le $X=(u_1^{(1)},u_2^{(1)},\tau_1^{(1)},\tau_2^{(1)})^{tr}$ be the leading order term of the solution. $X$ satisfies the differential equation
\begin{equation} \label{ODE_X}
X'=A X+g(t^*) \,,
\end{equation}
where $A$ and $g$ are defined as in \eqref{A} and \eqref{g} respectively. Notice that $g$ can be written as
\begin{equation}
g(t^*)=\sum_{j=1}^{+ \infty} \hat g_{s,j} \sin(j t^*)+\hat g_{c,j} \cos(j t^*) \,,
\end{equation}
where $\hat g_{s,j}=(0,0,J_s \tilde a_{1,j},J_s \tilde a_{2,j})^{tr}$ and $\hat g_{c,j}=(0,0,J_s \tilde b_{1,j},J_s \tilde b_{2,j})^{tr}$.
We are interested in computing the periodic orbit of the system, so we look for a solution in the following form
\begin{equation}
X(t^*)=\sum_{j=1}^{+ \infty} \hat X_{s,j} \sin(j t^*)+\hat X_{c,j} \cos(j t^*) \,,
\end{equation}
with $\hat X_{s,j}, \hat X_{c,j} \in \mathbb{R}^4$ for every $j$. It is not difficult to see that the ODE \eqref{ODE_X} is equivalent to the set of linear systems
\begin{equation} \label{linear_system_compact_appendix}
\begin{pmatrix}
-A& \ -j\mathbb{1}& \\
j\mathbb{1}& \ -A \\
\end{pmatrix} \left( \begin{array}{c} \hat X_{s,j} \\ \hat X_{c,j} \end{array} \right)= \left( \begin{array}{c} \hat g_{s,j} \\ \hat g_{c,j} \end{array} \right) \,,
\end{equation}
for $j \in \mathbb{N}$. Now suppose that 
\begin{equation}
\hat X_{s,j}=\left( \begin{array}{c} a_{1,j} \\ a_{2,j} \\c_{1,j} \\ c_{2,j} \end{array} \right) 
\end{equation}
and
\begin{equation}
\hat X_{c,j}=\left( \begin{array}{c} b_{1,j} \\ b_{2,j} \\ d_{1,j} \\ d_{2,j} \end{array} \right) \,.
\end{equation}
We may rewrite \eqref{linear_system_compact_appendix} as follows
\begin{align}
&j a_{1,j}=Q d_{1,j}+R d_{2,j} \label{linear_sys_1_appendix} \\
&j a_{2,j}=R d_{1,j}+Q d_{2,j}  \label{linear_sys_2_appendix}\\
&j b_{1,j}=-Qc_{1,j}-Rc_{2,j} \label{linear_sys_3_appendix} \\
&j b_{2,j}=-Rc_{1,j}-Q c_{2,j} \label{linear_sys_4_appendix} \\
&j c_{1,j}+(\tilde J-KQ) d_{1,j}-KR d_{2,j}-J_{p} K b_{1,j}=J_{s} \tilde b_{1,j} \\
&j c_{2,j}+(\tilde J-K Q) d_{2,j}-KR d_{1,j}-J_{p} K b_{2,j}=J_{s}\tilde b_{2,j} \\
&j d_{1,j}-(\tilde J-KQ)c_{1,j}+K R c_{2,j}+J_{p}K a_{1,j}=-J_{s} \tilde a_{1,j} \\
&j d_{2,j}-(\tilde J-KQ)c_{2,j}+KR c_{1,j}+J_{p}K a_{2,j}=-J_{s} \tilde a_{2,j} \,.
\end{align}
Equations \eqref{linear_sys_1_appendix}-\eqref{linear_sys_4_appendix} allow us to express $a_{1,j}$, $a_{2,j}$, $b_{1,j}$, and $b_{2,j}$ as linear combinations of $c_{1,j}$, $c_2{j}$, $d_{1,j}$, and $d_{2,j}$. If we plug the resulting expressions into the last four equations we obtain the linear problems
\begin{equation}
M(j) \left( \begin{array}{c} c_{1,j} \\ c_{2,j} \\ d_{1,j} \\ d_{2,j} \end{array} \right)=J_s \left( \begin{array}{c} \tilde b_{1,j} \\ \tilde b_{2,j} \\ -\tilde a_{1,j} \\ -\tilde a_{2,j} \end{array} \right) \,,
\end{equation}
where
\begin{equation}
M(j)=
\begin{pmatrix}
j+J_{p} K Q/j& \ J_{p}KR/j& \ \tilde J-KQ& \ -KR \\
J_{p}KR/j& \ j+J_{p}KQ/j& \ -KR& \ \tilde J-KQ \\
KQ-\tilde J& \ KR& \ j+J_{p}KQ/j& \ J_{p}KR/j \\
KR& \ KQ-\tilde J& \ J_{p}KR/j& \ j+J_{p}KQ/j \\
\end{pmatrix} \,.
\end{equation}
It is not difficult to check that $M(j)$ is invertible for every $j \in \mathbb{N}$: the computations are similar to the ones done in section \ref{asymptotic} for the matrix $M$. So we can conclude our computation with the help of the matrices $M(j)^{-1}$, $j \in \mathbb{N}$.

\subsection{TSS with one muscle-like arm and one passive elastic arm}

In this subsection we would like to generalize the results of subsection \ref{asymptotic_elastic} by considering a more general periodic function $\tilde \sigma$. We assume that $\tilde \sigma$ is a periodic function with period $2 \pi$. We expand this function in Fourier series
\begin{equation}
\tilde \sigma(t^*)=\sum_{j=1}^{+ \infty} \tilde a_j \sin(j t^*)+\tilde b_j \cos(j t^*) \,.
\end{equation}
Let $X=(u_1^{(1)},u_2^{(2)},\tau_2^{(2)})^{tr}$ be the leading order term of the solution. $X$ satisfies the ODE
\begin{equation} \label{ODE_X_elastic}
X'=B X+g(t^*) \,,
\end{equation}
where $B$ and $g$ are defined by equations \eqref{B} and \eqref{g_elastic} respectively. Notice that $g$ can be written as
\begin{equation}
g(t^*)=\sum_{j=1}^{+ \infty} \hat g_{s,j} \sin(j t^*)+\hat g_{c,j} \cos(j t^*) \,,
\end{equation}
where $\hat g_{s,j}=(0,0,J_s \tilde a_j)^{tr}$ and $\hat g_{c,j}=(0,0,J_s \tilde b_j)^{tr}$. We would like to compute the periodic orbit of the system, so we look for a solution in the form
\begin{equation}
X(t^*)=\sum_{j=1}^{+ \infty} \hat X_{s,j} \sin(j t^*)+\hat X_{c,j} \cos(j t^*) \,,
\end{equation}
with $\hat X_{s,j}, \hat X_{c,j} \in \mathbb{R}^3$ for every $j$. The ODE \eqref{ODE_X_elastic} is equivalent to the set of linear systems
\begin{equation} \label{linear_system_compact_elastic_appendix}
\begin{pmatrix}
-B& \ -j \mathbb{1}& \\
j \mathbb{1}& \ -B \\
\end{pmatrix} \left( \begin{array}{c} \hat X_{s,j} \\ \hat X_{c,j} \end{array} \right)=\left( \begin{array}{c} \hat g_{s,j} \\ \hat g_{c,j} \end{array} \right) \,,
\end{equation}
for $j \in \mathbb{N}$.
Now suppose that 
\begin{equation}
\hat X_{s,j}=\left( \begin{array}{c} a_{1,j} \\ a_{2,j} \\c_{j} \end{array} \right) 
\end{equation}
and
\begin{equation}
\hat X_{c,j}=\left( \begin{array}{c} b_{1,j} \\ b_{2,j} \\ d_{j} \end{array} \right) \,.
\end{equation}
Let us rewrite \eqref{linear_system_compact_elastic_appendix} explicitly
\begin{align}
&PHa_{1,j}+Rc_j+jb_{1,j}=0 \\
&RHa_{1,j}+Qc_j+jb_{2,j}=0 \\
&ja_{1,j}-PHb_{1,j}-Rd_j=0 \\
&ja_{2,j}-RHb_{1,j}-Qd_j=0 \\
&-RKHa_{1,j}-KJ_pa_{2,j}+(\tilde J-KQ)c_j-jd_j=J_s \tilde a_j \\
&jc_j-RKHb_{1,j}-KJ_pb_{2,j}+(\tilde J-KQ)d_j=J_s \tilde b_j \,.
\end{align}
Using the first four equations we can express $a_{1,j}$, $a_{2,j}$, $b_{1,j}$, and $b_{2,j}$ as linear combinations of $c_j$ and $d_j$. Then we can plug the resulting expressions into the last to equations. The result is the following linear problem
\begin{equation} \label{linear_problem_N}
N(j) \left( \begin{array}{c} c_j \\ d_j  \end{array} \right)= J_s \left( \begin{array}{c}  \tilde a_j\\ \tilde b_j  \end{array} \right) \,,
\end{equation}
where
\begin{equation}
N(j)=
\frac{1}{j^2+P^2H^2} 
\begin{pmatrix}
N_1(j) & \ -N_2(j) & \\
N_2(j) & \ N_1(j) & \\
\end{pmatrix} \,,
\end{equation}
and
\begin{align}
N_1(j)&=(J_p+J_s)(j^2+P^2H^2)+KH^2(PR^2-P^2Q)+KHJ_pR^2/j^2-J^2 KQ \\
N_2(j)&=j^3+j H^2P^2+j KHR^2+KH^2J_p(P^2Q-PR^2)+j^2 QKJ_p \,.
\end{align}
A simple computation shows that $\det(N(j))>0$ for every $j$. So with the help of the matrices $N(j)^{-1}$ we can compute the coefficients of our leading order approximation.

\section*{Acknowledgments} 
Support by the European Research Council through the ERC Advanced Grant 340685-MicroMotility is gratefully acknowledged.

\bibliography{biblioTSS}
\bibliographystyle{unsrt}

\end{document}